\documentclass[AMS,STIX1COL]{WileyNJD-v2}
\usepackage{moreverb}
\usepackage{bm}
\usepackage[caption=false]{subfig}
\usepackage{tikz}
\usetikzlibrary{spy}

\def\draftmode{false}

\newcommand{\wss}{\text{WSS}}
\newcommand{\tawss}{\text{TAWSS}}
\newcommand{\osi}{\text{OSI}}
\newcommand{\rrt}{\text{RRT}}

\newcommand\BibTeX{{\rmfamily B\kern-.05em \textsc{i\kern-.025em b}\kern-.08em
T\kern-.1667em\lower.7ex\hbox{E}\kern-.125emX}}

\newcommand{\review}[1]{\textcolor{black}{#1}}

\articletype{Article Type}%

\received{<day> <Month>, <year>}
\revised{<day> <Month>, <year>}
\accepted{<day> <Month>, <year>}

\begin{document}

\title{\textit{In silico} reproduction of the pathophysiology of in-stent restenosis}

\author[1]{Kiran Manjunatha$^*$}

\author[2]{Anna Ranno$^\#$}

\author[1]{Jianye Shi}

\author[3]{Nicole Schaaps}

\author[3]{Pakhwan Nilcham}

\author[3]{Anne Cornelissen}

\author[3]{Felix Vogt}

\author[2]{Marek Behr}

\author[1]{Stefanie Reese}

\authormark{Manjunatha \textsc{et al}}

\address[1]{\orgdiv{Institute of Applied Mechanics}, \orgname{RWTH Aachen University}, \orgaddress{\state{Aachen}, \country{Germany}}}

\address[2]{\orgdiv{Chair for Computational Analysis of Technical Systems}, \orgname{RWTH Aachen University}, \orgaddress{\state{Aachen}, \country{Germany}}}

\address[3]{\orgdiv{Department of Cardiology, Vascular Medicine and Intensive Care}, \orgname{RWTH Aachen University}, \orgaddress{\state{Aachen}, \country{Germany}}}

\corres{$^*$Kiran Manjunatha, Mies-van-der-Rohe-Str. 1, 52074 Aachen, Germany
.\\
\email{kiran.manjunatha@ifam.rwth-aachen.de}\\
$^{\#}$Anna Ranno, Schinkelstraße 2, 52062 Aachen, Germany.\\
\email{ranno@cats.rwth-aachen.de}}


\abstract[Abstract]{The occurrence of in-stent restenosis following percutaneous coronary intervention highlights the need for the creation of computational tools that can extract pathophysiological insights and optimize interventional procedures on a patient-specific basis. In light of this, a comprehensive framework encompassing multiple physical phenomena is introduced in this work. This framework effectively captures the intricate interplay of chemical, mechanical, and biological factors. In addition, computational approaches for the extraction of hemodynamic indicators that modulate the severity of the restenotic process are devised. Thus, this marks a significant stride towards facilitating computer-assisted clinical methodologies.}


\maketitle

\section{Introduction}
Due to the widespread adoption of an unhealthy lifestyle by a significant portion of the current populace, characterized by the consumption of highly processed foods, a lack of physical activity, and habits like smoking, the occurrence of cardiovascular ailments is exacerbated. A prominent example is atherosclerotic cardiovascular disease (ASCVD), where the accumulation of substantial plaque within the coronary arteries leads to restricted blood flow, potentially resulting in ischemic episodes and, in severe instances, myocardial infarction. As a result, ASCVD stands as one of the primary contributors to global mortality.

To address ASCVD symptoms and restore unobstructed blood flow in blocked coronary arteries, the most commonly employed method is percutaneous coronary intervention (PCI), accompanied by the insertion of scaffold-like stents. Nevertheless, it's important to note that the procedure carries inherent risks, notably the potential for two significant and unforeseen complications: (1) stent thrombosis (ST), and (2) in-stent restenosis (ISR). This particular study zeroes in on the latter issue, namely, ISR.

\subsection{Pathophysiology of ISR}

Friction between the PCI balloon or stent and the arterial wall, coupled with the arterial stretching that occurs during PCI, causes a partial or complete removal of the endothelial monolayer. This endothelial layer serves as a protective barrier, preventing the exposure of underlying contents to the bloodstream. However, the denudation of this layer results in the exposure of extracellular materials to the blood flow, initiating a series of inflammatory reactions. These responses encompass platelet aggregation, the release of cytokines and substances that promote cell division into the subendothelial area, prompting the migration and multiplication of smooth muscle cells (SMCs), and the infiltration of circulating monocytes.

Consequently, this sequence triggers unregulated tissue growth, culminating in the reoccurrence of blockages in the blood vessels. This outcome directly contradicts the purpose of PCI. This process is referred to as neointimal hyperplasia. 

Additionally, the perturbations in the hemodynamics within the coronary artery, brought about by stent-implantation, lead to low/oscillatory wall shear stress (WSS) on the vessel lining. This then leads to a dysfunctional behavior of the endothelium and consequently enhances the development of neointima \citep{nakazawa2008delayed, koskinas2012role, cecchi2011role, jenei2016wall}. 

Despite the substantial reduction in the occurrence of ISR achieved through the use of modern drug-eluting stents (DESs), the problem persists due to the individualized nature of patients' inflammatory responses and the limitations of the medical system in fine-tuning PCI procedures to match these responses appropriately.
 
\subsection{State of the art computational models}

Numerous computational frameworks have been proposed to simulate ISR. These models can be broadly categorized as follows:

\begin{enumerate}
    \item [(a)] Approaches that combine cellular automata (CA) with agent-based modeling (ABM), as evidenced in references \cite{evans2008}, \cite{tahir2011}, \cite{Zahedmanesh2014AMM}, and \cite{Zun2021EffectsOL}
    \item [(b)] Phenomenological continuum mechanical models, as demonstrated in the works \cite{Garikipati2004ACT}, \cite{fereidoo2017}, and \cite{he2020}.
    \item [(c)] Multiscale multiphysics-based chemo-mechano-biological continuum models, elaborated upon in references \cite{budu2008} and \cite{escuer2019}.
\end{enumerate}

However, computational models that incorporate the impact of drug pharmacokinetics and pharmacodynamics, particularly in the context of drugs integrated within drug-eluting stents (DESs), have been relatively limited in the literature. A handful of works, \cite{Caiazzo2011ACA}, \cite{Rossi2012BioresorbablePC}, and \cite{MCQUEEN2022992} to name a few, have delved into these aspects.

Another area of current research is the choice of a constitutive model for blood, which is relevant in the context of a fully coupled fluid-structure interaction (FSI) model for ISR. In large arteries with physiological flow, blood can be approximated as a homogeneous Newtonian fluid without altering the overall flow characteristics. For capillary arteries or in pathological conditions, shear thinning effects and blood damage cannot be neglected \citep{behr2006models, leuprecht2001computer, gijsen1999influence, behbahani2009review,hassler2019variational, guglietta2020effects, Sasse}. One of the most relevant blood flow indicators to assess the well-being of the artery wall is WSS, and hence an FSI framework that incorporates the effects of WSS on the pathophysiology of ISR is of utmost interest.

The current study therefore contributes to the advancement of computational methodologies for modeling ISR. It specifically focuses on the intricate interplay between crucial mediators and the kinetics of drug elution, providing a substantial resolution that can yield valuable insights and facilitate the customization of interventional procedures. Additionally, to aid the development of an FSI framework, hemodynamics aspects of stent implantation and the methodology for extraction of key blood flow indicators are established. \\

 \subsection{Overview}
The authors herein present two constituents that make up a fully resolved FSI framework for modeling ISR. This includes the development of a multiphysics model for the arterial wall that captures the interactions between significant mediators of ISR in combination with a hemodynamic computation setup that derives key blood flow quantities that influence the restenotic process.

\section{Model description} 
\subsection{Arterial wall model}
\subsubsection{Continuum mechanical modeling}
The arterial wall is assumed to be composed of two layers, the media and the adventitia, each consisting of two helices of collagen fibres embedded in an isotropic ground matrix. To model the volumetric growth involved in the restenotic process, we adopt the well-known multiplicative split of the deformation gradient $\bm{F}$ \cite{rodriguez1994}, i.e.,
\begin{equation}\label{mult_split}
    \boldsymbol{F} = \boldsymbol{F}_e\,\boldsymbol{F}_g,
\end{equation}
where $\bm{F}_g$ maps the referential geometry to an intermediate stress-free grown state, while $\bm{F}_e$ achieves the compatibility of deformations. Upon further decomposition of the growth part $\bm{F}_g$ into a pure stretch part and a pure rotational part \cite{HOLTHUSEN2023105174}, i.e.,
\begin{equation}
    \boldsymbol{F}_g = \boldsymbol{R}_g\,\boldsymbol{U}_g,
\end{equation}
we can rewrite the decomposition in Eq. \ref{mult_split} as
\begin{equation}
    \boldsymbol{F} = \underbrace{\bm{F}_e\,\bm{R}_g}_{:=\bm{F}_{*}}\,\bm{U}_g = \bm{F}_{*}\,\bm{U}_g.
\end{equation}
The right Cauchy-Green tensor associated with the mapping $\bm{F}_*$ shall then be
\begin{equation}\label{rcg_*}
\boldsymbol{C}_* = \bm{F}^T_*\,\bm{F}_* = \boldsymbol{U}_g^{-1}\, \boldsymbol{C}\, \boldsymbol{U}_g^{-1}.
\end{equation}
The Helmholtz free energy per unit volume in the reference configuration is additively split into an isotropic part associated with the isotropic ground matrix, and an anisotropic part corresponding to the collagen fibres, i.e.,
\begin{equation} \label{hfe}
\psi = \psi_{\sf iso} + \psi_{\sf ani}. 
\end{equation}
With $\bm{U}_g$ as an internal variable, the Helmholtz free energy for the isotropic ground matrix, encompassing the behaviors of the smooth muscle cells, elastins, and proteoglycans, is prescribed to be of the Neo-Hookean form as
\begin{equation}\label{iso_hfe}
\psi_{\sf iso} (\bm{C},\bm{U}_g) = \displaystyle{\frac{\mu}{2}}\left(\text{tr}\,\boldsymbol{C}_* - 3\right) - \mu\,\text{ln}\,J_* + \displaystyle{\frac{\Lambda}{4}}\left(J_*^2 - 1 - 2\,\text{ln}\,J_* \right),
\end{equation}
where $J_* = \sqrt{{\sf det}(\bm{C}_*)}$. We herein embed the directionality of the helices of collagen fibres via the generalized structural tensor approach \cite{gasser2006}, defining them to be of the form
\begin{eqnarray}\label{struct_tens}
\boldsymbol{H}_i &=& \kappa\,\boldsymbol{I} + \left(1 - 3\,\kappa \right)\,\boldsymbol{a}_{0i} \otimes \boldsymbol{a}_{0i}
\end{eqnarray}
for each helix ($i = 1, 2$). Here $\kappa$ refers to the dispersion parameter, and $\bm{a}_{0i}$ to the collagen orientations in the reference configuration. The Helmholtz free energy associated with the collagen fibres is then chosen to be of the form
\begin{equation}\label{aniso_hfe}
\psi_{\sf ani} (\boldsymbol{C}, \boldsymbol{H}_1, \boldsymbol{H}_2, c^0_{{}_C}) = \displaystyle{\frac{k_1}{2k_2}}\sum_{i=1,2} \left(\text{\sf exp}\left[k_2\langle E_i\rangle ^2 \right]-1\right),
\end{equation}
where 
\begin{equation}\label{GL_strain}
    E_i := \boldsymbol{H}_i : \boldsymbol{C} - 1,
\end{equation}
and the stress-like parameter $k_{1}$ is scaled according to the local collagen concentration in the reference configuration ($c^0_{{}_C}$), i.e.,
\begin{equation}
    k_1 = \bar{k}_1 \,\displaystyle{\left(\frac{c^{0}_{{}_C}}{c_{{}_{C,eq}}}\right)},
\end{equation}
where $\bar{k}_1$ is the stress-like material parameter for healthy collagen and $c_{{}_{C,eq}}$ is the homeostatic collagen concentration in a healthy artery. 

Volumetric growth is achieved via the direct prescription of the growth stretch tensor $\bm{U}_g$ \cite{Lubarda2002OnTM} based on the amount of dispersion $\kappa$, which can be histologically inferred. Two forms are proposed herein, namely,
\begin{eqnarray}
(\kappa \approx 0)&:&\quad  \bm{U}_g:= \bm{I} + (\vartheta - 1)\,\bm{\gamma}\otimes\bm{\gamma},\quad \bm{\gamma}:= \displaystyle{\frac{\bm{a}_{01}\times\bm{a}_{02}}{||\bm{a}_{01}\times\bm{a}_{02}||}}, \quad \vartheta:= \displaystyle{\frac{\rho^0_{{}_{S}}}{\rho_{{}_{S,eq}}}}\nonumber\\  
(\kappa > 0)&:& \quad \bm{U}_g:= \vartheta\,\bm{I}, \quad \vartheta:= \displaystyle{\left(\frac{\rho^0_{{}_{S}}}{\rho_{{}_{S,eq}}}\right)^{1/3}},
\end{eqnarray}
where $\rho^0_{{}_{S}}$ is the local SMC density in the reference configuration, and $\rho^0_{{}_{S,eq}}$ is the SMC density in a healthy homeostatic vessel.

\subsubsection{Balance equations for mediators}
The balance equations for the concentrations of PDGF ($c_{{}_P}$), TGF-$\beta$ ($c_{{}_T}$), ECM with collagen as the primary constituent ($c_{{}_C}$), and the drug ($c_{{}_D}$), in addition to the densities of SMCs ($\rho_{{}_S}$) and ECs ($\rho_{{}_E}$) are set up to model the chemo-biological interactions involved in neointimal hyperplasia (See \cite{manjunatha2023} for details). The advection-reaction-diffusion equation forms the basis for the aforementioned balance of mediators, the general form of which in the Eulerian description is
\begin{equation}\label{ard_eq_general_form_eul}
    \underset{\lower.5em \hbox{\text{rate}}}{\displaystyle{\left.\frac{\partial \phi}{\partial t}\right|_{\bm{x}}}} + \underbrace{\text{\sf div} \left(\phi\,\boldsymbol{v}\right)}_{\lower.3em \hbox{\text{advection}}} = \underbrace{\text{\sf div} \left(k\,\text{\sf grad} \phi\right)}_{\lower.35em \hbox{\text{diffusion}}} + \underbrace{\overset{\text{source}}{\mathcal{R}} - \overset{\text{sink}}{\mathcal{S}}}_{\lower.4em \hbox{\text{reaction}}}.
\end{equation}
The particularized balance equations for the mediators are set up based on the physiological response of a stented artery in an inflammatory setting and are summarized to be 
\begin{alignat}{15}
    &\textbf{PDGF}: &&\nonumber\\
    &\displaystyle{\left.\frac{\partial c_{{}_P}}{\partial t}\right|_{\bm{x}}} +&& \text{\sf div} \left(c_{{}_P}\,\boldsymbol{v}\right) = &&\underbrace{\text{\sf div} \left(D_{{}_{P}}\,\text{\sf grad}\,c_{{}_{P}}\right)}_{\text{diffusion}} + \underbrace{\left((1 - r_{{}_{\eta}}) + r_{{}_{\eta}}\, f_{{}_{P1}}(c_{{}_D})\right)\eta_{{}_P} \rho_{{}_{S}}\, c_{{}_{T}}}_{\substack{\text{secretion by SMCs}\\\text{and macrophages}}} - \underbrace{\varepsilon_{{}_P} \,f_{{}_{P2}}(c_{{}_T})\,\rho_{{}_{S}}\, c_{{}_{P}}}_{\substack{\text{receptor}\\\text{internalization}}},\nonumber\\
    \\
    &\textbf{TGF-}\bm{\beta}:&&\nonumber\\
    &\displaystyle{\left.\frac{\partial c_{{}_T}}{\partial t}\right|_{\bm{x}}} +&& \text{\sf div} \left(c_{{}_T}\,\boldsymbol{v}\right) = &&\underbrace{\text{\sf div} \left(D_{{}_{T}}\,\text{\sf grad}\,c_{{}_{T}}\right)}_{\text{diffusion}} - \underbrace{\varepsilon_{{}_T} \,\rho_{{}_{S}}\, c_{{}_{T}}}_{\substack{\text{receptor}\\\text{internalization}}},\nonumber\\
    \\
    &\textbf{ECM}:&&\nonumber\\
    &\displaystyle{\left.\frac{\partial c_{{}_C}}{\partial t}\right|_{\bm{x}}} +&& \text{\sf div} \left(c_{{}_C}\,\boldsymbol{v}\right) = &&\underbrace{\eta_{{}_C} \rho_{{}_{S}} \left(1 - \displaystyle{\frac{c_{{}_{C}}}{c_{{}_{C,th}}}}\right)}_{\substack{\text{secretion by}\\\text{synthetic SMCs}}} - \underbrace{\varepsilon_{{}_C} \, c_{{}_{P}}\,c_{{}_{C}}}_{\substack{\text{MMP-induced}\\\text{degradation}}},\nonumber\\
    \\
    &\textbf{drug}&& \textbf{(rapamycin-analogs):}\nonumber\\
    &\displaystyle{\left.\frac{\partial c_{{}_D}}{\partial t}\right|_{\bm{x}}} +&& \text{\sf div} \left(c_{{}_D}\,\boldsymbol{v}\right) = &&\underbrace{\text{\sf div} \left(D_{{}_{D}}\,\text{\sf grad}\,c_{{}_{D}}\right)}_{\text{diffusion}} - \underbrace{\varepsilon_{{}_{D1}} \,\rho_{{}_{S}}\, c_{{}_{D}}}_{\substack{\text{receptor}\\\text{internalization}}},\nonumber\\
    \\
    &\textbf{EC}:&&\nonumber\\
     &\displaystyle{\left.\frac{\partial \rho_{{}_{E}}}{\partial t}\right|_{\bm{x}}} +&& \text{\sf div}_{{}_{\Gamma}} \left(\rho_{{}_{E}}\,\boldsymbol{v}_{{}_{\Gamma}}\right) = &&\underbrace{\text{\sf div}_{{}_{\Gamma}} \left(D_{{}_{E}}\,\text{\sf grad}_{{}_{\Gamma}}\,\rho_{{}_{E}}\right)}_{\text{diffusion}}
+ \underbrace{\eta_{{}_{E}}\, f_{{}_{E1}}(c_{{}_D})\,\rho_{{}_{E}}\,\left(1 - \frac{\rho_{{}_{E}}}{\rho_{{}_{E,eq}}}\right)}_{\text{proliferation}} - \underbrace{\varepsilon_{{}_{E}}\,f_{{}_{E2}}(c_{{}_D})\,\rho_{{}_{E}}}_{\text{apoptosis}},\nonumber\\
 \end{alignat}
 \begin{alignat}{5}
    &\textbf{SMC}:&& \nonumber\\
    &\displaystyle{\left.\frac{\partial \rho_{{}_S}}{\partial t}\right|_{\bm{x}}} +&& \text{\sf div} \left(\rho_{{}_S}\,\boldsymbol{v}\right) = &&
      - \underbrace{\text{\sf div}\left(\rho_{{}_S}\,\overbrace{\chi_{{}_{S1}} \left(1 - \displaystyle{\frac{c_{{}_{C}}}{c_{{}_{C,th}}}}\right)\,\text{\sf grad}\,c_{{}_{P}}}^{ \bm{v}_{{}_{S1}}} \right)}_{\text{chemotaxis}} +  \underbrace{ \text{\sf div}\left(\rho_{{}_S}\,\overbrace{\chi_{{}_{S2}} \,f_{{}_{S1}}(c_{{}_P}) \,\text{\sf grad}\,c_{{}_{C}}}^{- \bm{v}_{{}_{S2}}}\right)}_{\text{haptotaxis}} &\nonumber\\
     & && && + \underbrace{\eta_{{}_S}\, f_{{}_{S2}}(c_{{}_P})\,f_{{}_{S3}}(c_{{}_T})\,f_{{}_{S4}}(c_{{}_D})\, \rho_{{}_{S}} \left(1 - \displaystyle{\frac{c_{{}_{C}}}{c_{{}_{C,th}}}}\right)}_{\text{proliferation}},\nonumber\\
\end{alignat}
the scaling functions embedded in the above balance equations being defined to be
\begin{eqnarray}
    f_{{}_{P1}}(c_{{}_D}) &:=& \displaystyle{\exp ({-l_{{}_{P1}}\,c_{{}_D}})} \in [0,1],\\
    \nonumber\\
    f_{{}_{P2}}(c_{{}_T}) &:=& \frac{1}{1 + \exp\left({l_{{}_{P2}}\left(c_{{}_T} - c_{{}_{T,th}}\right)}\right)}\,\,\,\in [0,1]\\
    \nonumber\\
    f_{{}_{S1}}(c_{{}_P}) &:=& \frac{1}{1 + \exp\left({-l_{{}_{S1}}\left(c_{{}_P} - c_{{}_{P,th}}\right)}\right)}\,\,\,\in [0,1]\\
    \nonumber\\
    f_{{}_{S2}}(c_{{}_P}) &:=& 1 - \exp(-l_{{}_{S2}}\,c_{{}_P}) \in [0,1],\\
    \nonumber\\
    f_{{}_{S3}}(c_{{}_T}) &:=& \frac{1}{1 + \exp\left({l_{{}_{S3}}\left(c_{{}_T} - c_{{}_{T,th}}\right)}\right)}\,\,\,\in [0,1],\\
    \nonumber\\
    f_{{}_{S4}}(c_{{}_D}) &:=& 1 - \frac{1}{100}\,\left(\frac{A_{{}_S}\,c^{\alpha}_{{}_D}}{c^{\alpha}_{{}_D} + B_{{}_S}^{\alpha}}\right)\,\,\,\in [0,1],\\
    f_{{}_{E1}}(c_{{}_D}) &:=& 1 - \frac{1}{100}\,\left(\frac{A_{{}_E}\,c^{\beta}_{{}_D}}{c^{\beta}_{{}_D} + B_{{}_E}^{\beta}}\right)\,\,\,\in [0,1],\\
    \nonumber\\
    f_{{}_{E2}}(c_{{}D}) &:=& 1 - \exp(-l_{{}_E}\,c_{{}_D}) \in [0, 1],
\end{eqnarray}
and $\bm{v}_{{}_{S1}}$ and $\bm{v}_{{}_{S2}}$ referring to the chemotactic and haptotactic velocities respectively. It is to be mentioned here that all the equations for the mediators except that for the ECs are solved in the bulk domain of the arterial wall, while that for the ECs is solved only on the luminal surface. Hence the usage of the surface gradient and divergence operators in the balance equation for EC density. We herein employ the Lagrangian equivalent of the general form in Eq. \ref{ard_eq_general_form_eul}, which is obtained from the transformation of the terms involved, given by
\begin{equation}\label{ard_eq_general_form_lag}
   J^{-1}\,\dot{{\phi}^{0}} = \displaystyle{J^{-1}{\sf Div}\,\left[k\,\bm{C}^{-1}\,\left(\text{{\sf Grad}}(\phi^0) - \left(\frac{\phi^0}{J}\right)\,{\sf Grad}\, J\right)\right]} + \mathcal{R}_0 - \mathcal{S}_0.
\end{equation}
The quasi-static balance of linear momentum 
\begin{equation}\label{mom_bal}
{\sf Div}\, (\bm{F}\,\bm{S}) + \boldsymbol{B} = \boldsymbol{0},
\end{equation}
forms the basis for modeling the structural response of the arterial wall, $\bm{S}$ being the second Piola-Kirchhoff stress tensor, and $\bm{B}$ the body force vector in the reference configuration. The inertial effects of the added mass that results in the slow growth process are hence ignored.

\subsection{Blood flow model}
Blood flow is modeled by means of incompressible Navier-Stokes equations and a Newtonian constitutive model. At this validation stage, walls are assumed to be rigid. \textit{No-slip} boundary conditions are imposed on the artery wall $\Gamma_{w}$ and on the stent inner surface $\Gamma_{stent}$. On $\Gamma_{out}$, we impose for the velocity to be perfectly orthogonal to the outflow surface.
\begin{figure} [htbp!]
    \centering
    \subfloat[Color-coded boundaries: $\Gamma_{in}$ in yellow, $\Gamma_w$ in grey, $\Gamma_{stent}$ in pink and $\Gamma_{out}$ in blue.]{
	   \resizebox*{!}{4.5cm}{\includegraphics[draft=\draftmode]{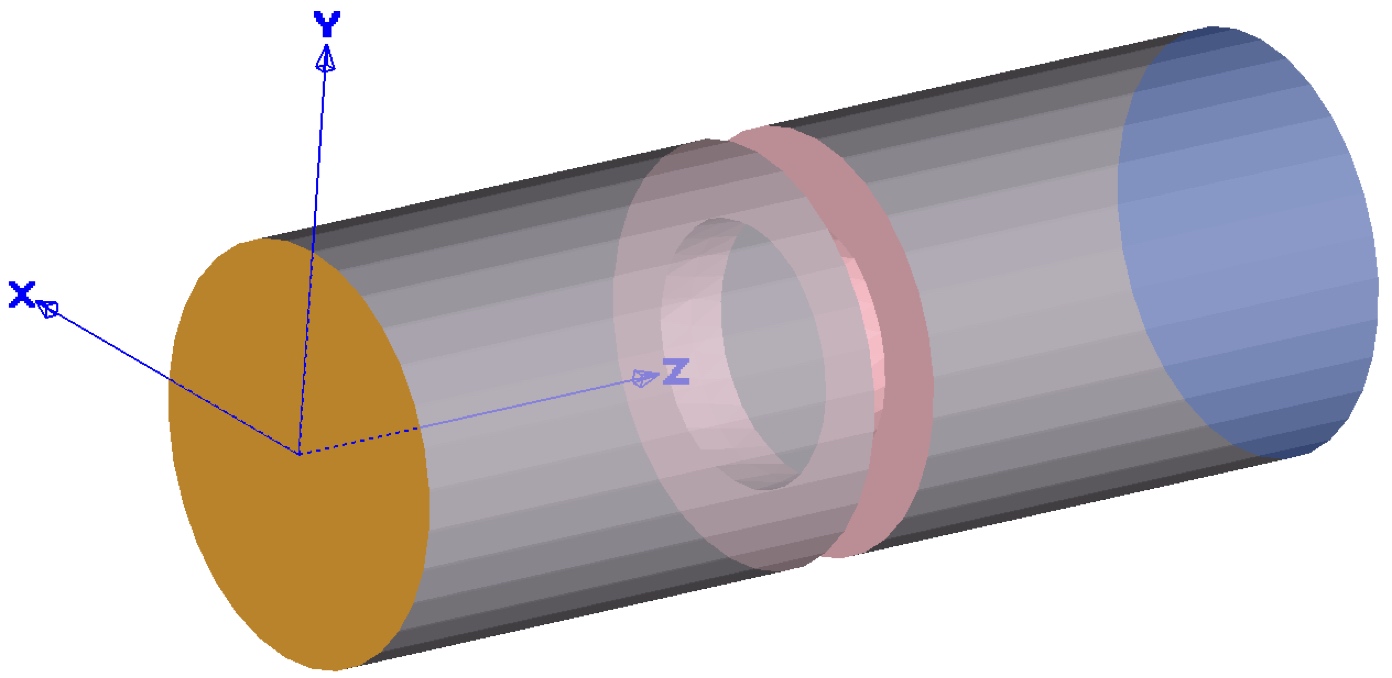}}\label{fig:boundaries}}
	\hspace{3pt}
    \subfloat[Flow rate over one heartbeat imposed at inflow boundary $\Gamma_{in}$.]{
	   \resizebox*{!}{4.5cm}{\includegraphics[draft=\draftmode]{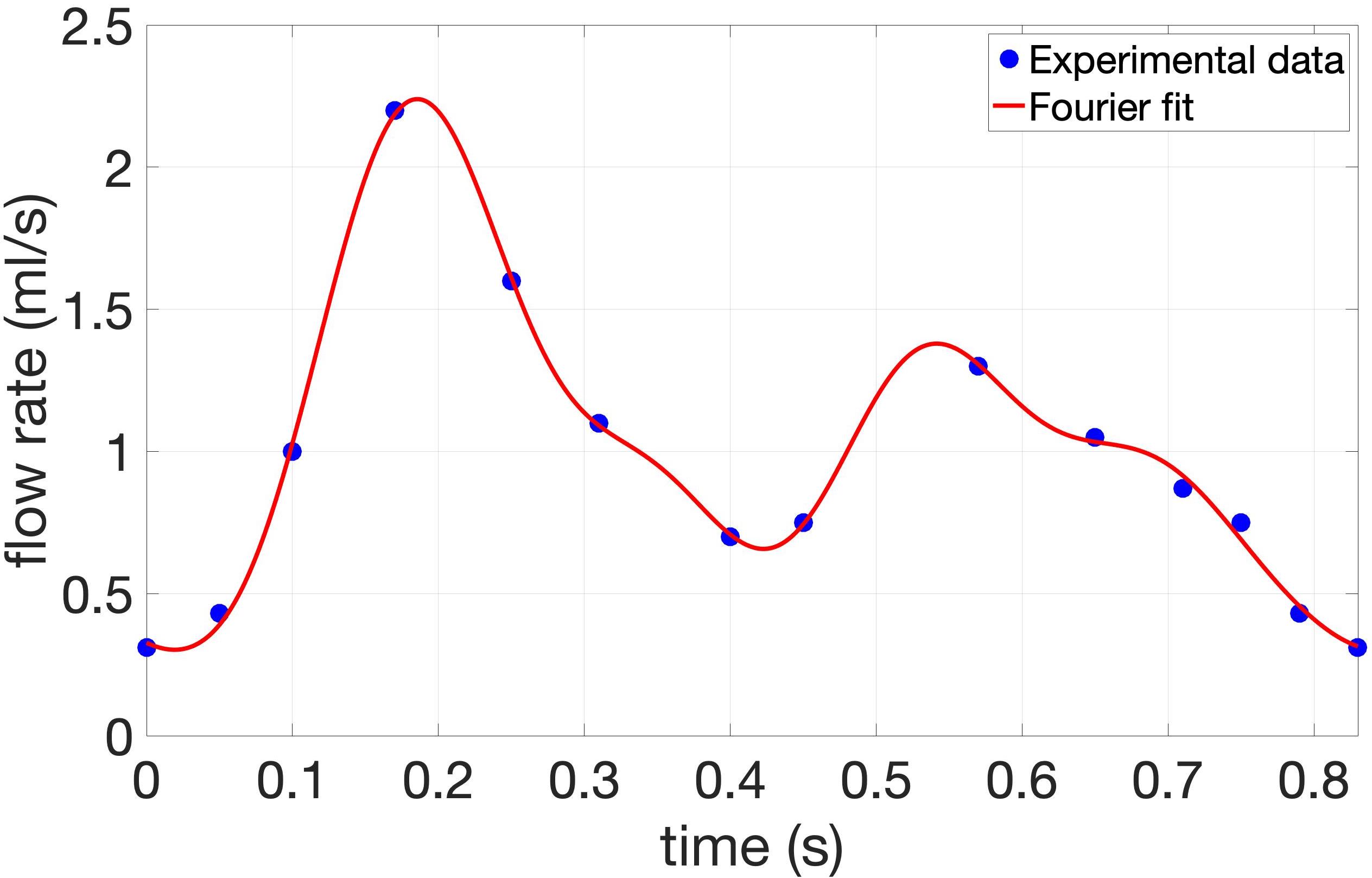}}\label{fig:flowRate}}
	\caption{Definition of boundaries and inflow boundary conditions.}\label{fig:BC}
\end{figure} The boundaries are depicted in Fig.\ref{fig:boundaries}. The inflow velocity imposed on $\Gamma_{in}$ is given by the function $\mathbf{g}$ as:
\begin{align}
	\mathbf{g}(\mathbf{x},t) & = \review{C(A(t))}Q(t)\left(1-\frac{|\mathbf{x}|^2}{r_0^2}\right)\mathbf{n} && \text { on } \Gamma_{in},
\end{align}
where \review{$C(A(t))$ is a function} used to convert the flow rate $Q(t)$ from ml/s to mm/s depending on area \review{$A(t)$} of the inlet surface $\Gamma_{in}$ and $r_0 = 1.8$ mm is the radius of the artery. \review{The function $C$ is derived from the assumption that for parabolic profile, the peak velocity $v_{\text{max}} = 2\bar{v} = 2 \frac{Q}{A}$ where $\bar{v}$ is the average velocity and the flow rate $Q$ and area $A$ are evaluated at fixed time $t$. In general, the area $A(t)$ can change in time and the flow rate is given in $\frac{ml}{s}$, hence
\begin{equation}
    C(A(t)) = \frac{2}{A(t)}C_{ml}^{mm^3}, \hspace{20pt} \text{and} \hspace{20pt} C_{ml}^{mm^3} = 1 \frac{ml}{s} = 10^3 \frac{mm^3}{s}.
\end{equation}
} In the case of rigid wall, the \review{circular inlet area $A(t)=A = \pi r_0^2$ is constant in time and the function $C(A)$ needs to be computed only once, independently of the flow rate of choice}. \review{Figure \ref{fig:flowRate} shows the flow rate $Q(t)$ over one heart beat in the right coronary artery based on experimental values from \citep{bertolotti2001numerical, hsiao2012hemodynamic}. In particular, the blue dots are extrapolated with WebPlotDigitizer \citep{Rohatgi2022} from Figure 7b in \citep{hsiao2012hemodynamic}, except for the minimum and maximum flow rate which are the only precise numerical values available. Hence, we specified a periodic minimum flow rate $Q(0 \hspace{1pt} s) = Q(0.83 \hspace{1pt} s) = 0.31 \frac{ml}{s}$ and the maximum flow rate $Q(0.17 \hspace{1pt} s) = 2.2 \frac{ml}{s}$, for a total of 14 evaluation points (see table \ref{interpData}).
\begin{center}
\begin{table}[htbp!]
\caption{Experimental data extrapolated from \citep{bertolotti2001numerical,hsiao2012hemodynamic}.}
\centering
\begin{tabular}{lccccccccccccccc}\toprule
    time $t$ [s] & 0 & 0.05 & 0.1 & 0.17 & 0.25 & 0.31 & 0.4 & 0.45 & 0.57 & 0.65 & 0.71, & 0.75 & 0.79 & 0.83\\\midrule
    $Q(t)$ [ml/s] & 0.31 & 0.43 & 1.0 & 2.2 & 1.6 & 1.1 & 0.7 & 0.75 & 1.3 & 1.05 & 0.87 & 0.75 & 0.43 & 0.31\\
    \bottomrule
\end{tabular}
\label{interpData}
\end{table}
\end{center}}
We interpolate the experimental data by means of Fourier series using MATLAB Curve Fitting Toolbox \citep{matlabcftool}. In the plot, the red line is obtained with 5 Fourier modes:
\begin{equation}
	Q(t)=\frac{a_0}{2}+\sum_{k=1}^{5}\left[a_k\cos\left(k\frac{2\pi}{T} \review{t}\right)+b_k\sin\left(k\frac{2\pi}{T}\review{t}\right)\right],
\end{equation}
where the coefficients $a_i$ and $b_i$ can be found in table \ref{Fouriercoeff} and $T=0.83$ s. A Fourier interpolation has multiple advantages: we can choose \review{the period $T$ to} maintain the physiological periodicity of heart beats, we can tune the interpolation accuracy choosing more Fourier modes and we impose a very smooth inflow velocity, which prevents possible instabilities due to the inflow boundary imposition.
We use GLS-stabilized FEM with $\mathbb{P}_1 - \mathbb{P}_1$ element pair in space \citep{donea2003finite} and BDF2 multi-step method in time \citep{brenan1995numerical}. We linearize the convective term by means of Newton-Raphson method and solve the resulting linear system with a GMRES solver \citep{saad1986gmres} and ILUT preconditioning \citep{saad2003iterative}.\\

\begin{center}
\begin{table}[htbp!]
\caption{Fourier coefficients of flow rate for each $k$-th mode.}
\centering
\begin{tabular}{lccccccc}\toprule
     & 0 & 1 & 2 & 3 & 4 & 5\\\midrule
    $a_k$ & 2.146 & \review{-}0.1646 & \review{-}0.5616 & \review{-}0.0658 & \review{-}0.0129 & 0.0592\\
    $b_k$ & - & 0.2692 & 0.1069 &  \review{-}0.1992 & \review{-}0.1287 & 0.0633\\
    \bottomrule
\end{tabular}
\label{Fouriercoeff}
\end{table}
\end{center}

The WSS is computed as:
\begin{equation}
	\wss = |\bm{\tau}| = \mathbf{t} \cdot (\bm{\sigma}\mathbf{n}),
\end{equation}
where $ \mathbf{t} $ is the tangent vector, $\mathbf{n}$ is the outward normal vector and $\bm{\sigma}$  is the Cauchy stress tensor. The stress tensor for incompressible and viscous flows is defined as:
\begin{equation}
	\bm{\sigma} = - p \mathbf{I} + 2 \mu \mathbf{E},
	\label{sigma}
\end{equation}
where $p=p(\mathbf{x},t)$ is the pressure, $\mu$ is the dynamic viscosity, $\mathbf{E}(\mathbf{u}) = \frac{1}{2}\left( \nabla \mathbf{u} + \nabla \mathbf{u}^\top\right)$ is the rate of strain tensor and $\mathbf{u}=\mathbf{u} ( \mathbf{x},t )$ is the velocity vector.
We compare two possible computations of WSS, that we refer to as $\bm{\tau}_1$ and $\bm{\tau}_2$ \citep{john2017influence}.
In the first case, we compute the WSS magnitude as the shear rate multiplied by the blood viscosity $\mu$.
The shear rate is given by:
\begin{equation}
		\dot{\gamma} = \sqrt{- 4 \textit{II}_{\mathbf{E}}} = \sqrt{2 \mathbf{E}\colon\mathbf{E}} = \sqrt{2 tr(\mathbf{E}^2)},
		\label{shearRate}
\end{equation}
where $\textit{II}_{\mathbf{E}}$ is the second invariant of the tensor $\mathbf{E}$ and $ tr(\mathbf{E}^2)$ is the trace and first invariant of the square tensor.
This simplification is derived from the assumption that for parabolic unidirectional flows, the tangent vector $ \mathbf{t} = \mathbf{t}_{blood} $ is uniquely defined, and $tr(E) =0$. Thus, the WSS tensor $\bm{\tau}_1$ reduces to:
\begin{equation}
	\wss_1 = |\bm{\tau}_1| = \mathbf{t} \cdot (\bm{\sigma}\mathbf{n}) = |2\mu \mathbf{E}\mathbf{n}|  = \mu \dot{\gamma},
\end{equation}
where $ \bm{\sigma}$ is defined in \eqref{sigma}.
If no assumptions on the flow are made, the WSS tensor $\bm{\tau}_2$ has the following expression:
\begin{equation}
	\bm{\tau}_2 = \bm{\sigma}\mathbf{n} - [(\bm{\sigma}\mathbf{n})\cdot \mathbf{n}]\mathbf{n} = 2 \mu (\mathbf{E}\mathbf{n} - [(\mathbf{E}\mathbf{n})\cdot \mathbf{n}]\mathbf{n}),
\end{equation}
and its magnitude is defined as $\wss_2 = |\bm{\tau}_2 |$.

\section{Results}

\subsection{Evolution of restenosis in a coronary arterial wall implanted with a XIENCE-V stent}

The weak forms of the equations established in the preceding section in the Lagrangian coordinates are discretized spatially using trilinear hexahedral elements for the bulk of the arterial wall, and bilinear quadrilateral elements for the luminal surface. The temporal discretization is achieved using the backward-Euler method. This entire framework is then put into action using the finite element software package FEAP, utilizing user-defined elements. The necessary parameters are sourced from existing literature whenever possible. In cases where such data is unavailable, choices are made that lead to physiological macroscopic results.

In order to assess the viability of the modeling framework within a complex geometric context, a section of the human coronary artery containing an implanted XIENCE-V stent (manufactured by Abbott Vascular Inc.) is taken into consideration. The region denoted as $\Gamma^d_{{}_E}$ signifies the area where the endothelium is assumed to have been completely removed due to the stent implantation procedure. Thus, the initial density $\rho_{{}_E}$ of endothelial cells (ECs) is set to zero within this region. Along the circular boundaries of the luminal surface ($\Gamma^h_{{}_E}$), the EC density is initialized to match the healthy equilibrium value, initiating the process of endothelial recovery. $\Gamma_{{}_{SS}}$ designates the segment where the stent makes contact with the luminal side, and here the influx of drugs is predefined. The extremities in the longitudinal direction are restricted from experiencing longitudinal displacements, while the cylindrical surfaces on the non-luminal side are constrained against radial movements. The stent apposition area remains fixed against all forms of displacement.

\begin{figure}[htbp!]
    \centering
    \includegraphics[scale=0.45]{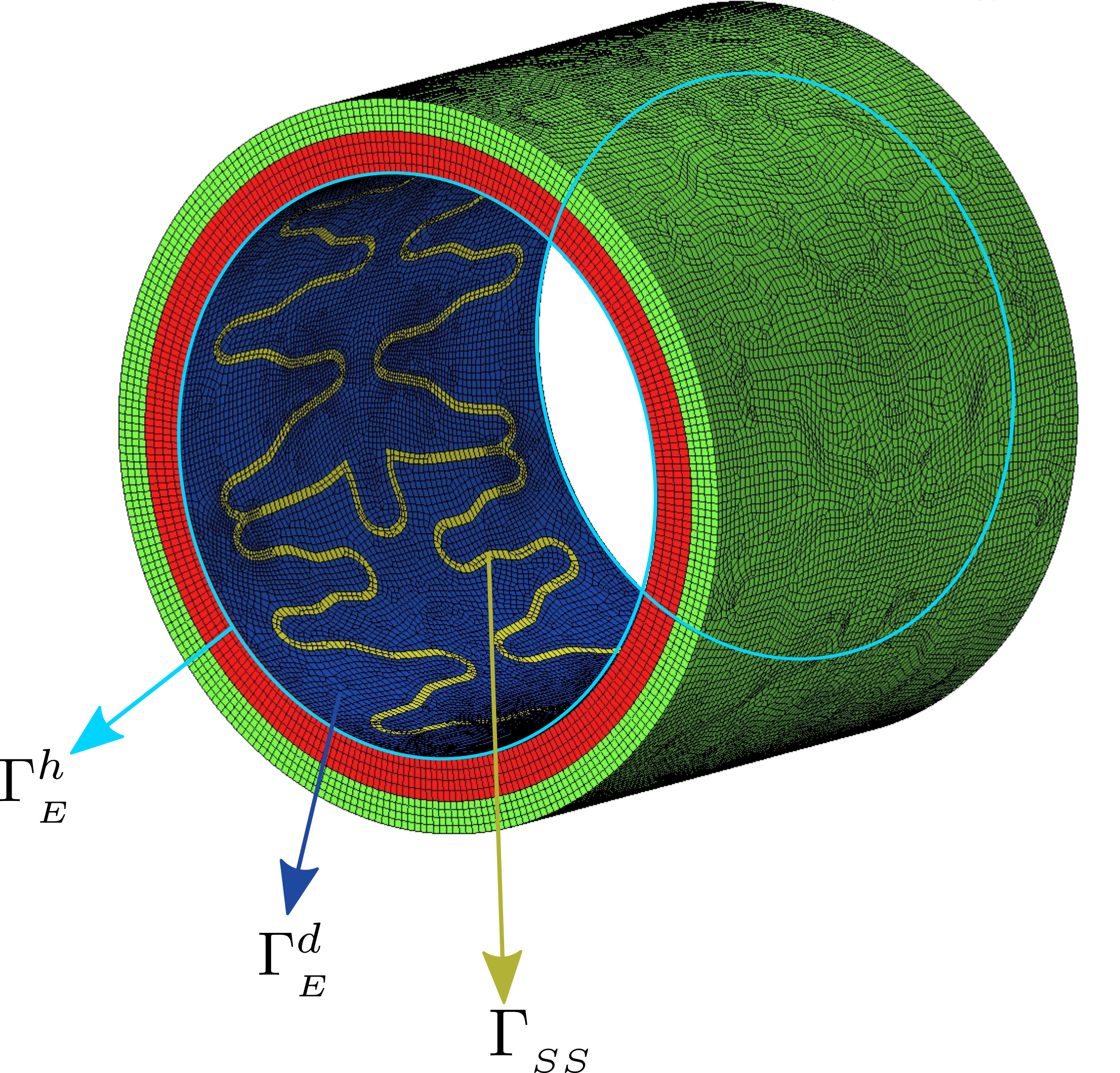}
    \caption{{XIENCE-V stent in a coronary artery.}}
    \label{fig_bvp_IFAM}
\end{figure}

\begin{figure}[htbp!]
    \centering
    \includegraphics[scale=0.7]{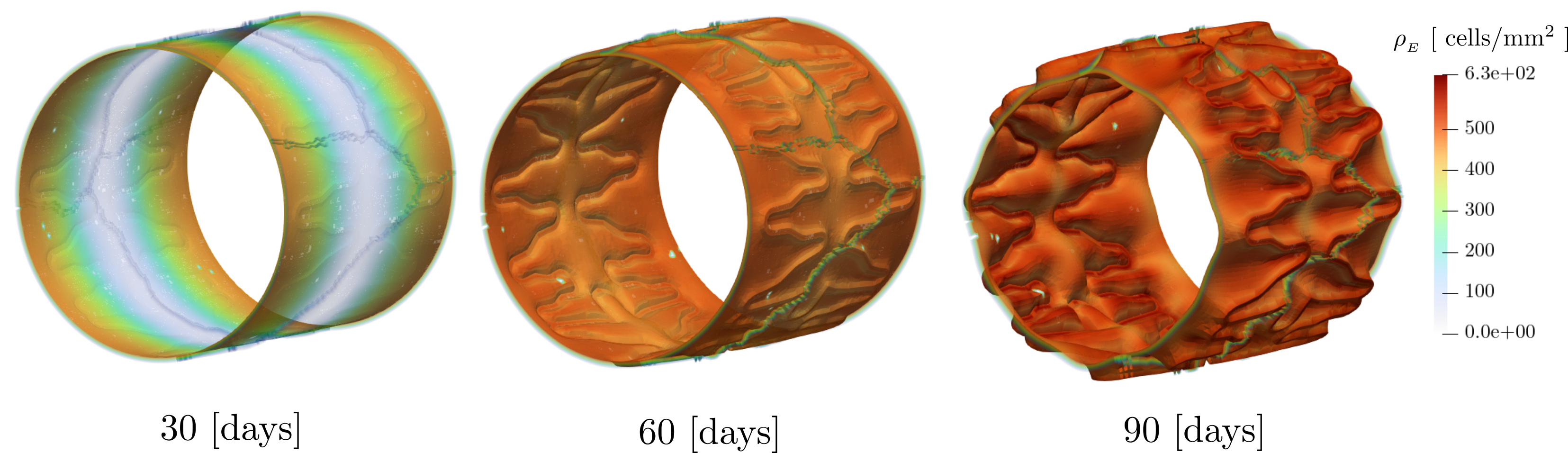}
    \caption{{Endothelial recovery in the case of a bare metal stent implantation.}}
    \label{fig_EC_healing_IFAM}
\end{figure}

The established model is employed to simulate a period of 90 [days]. Fig. \ref{fig_EC_healing_IFAM} illustrates the process of endothelial recovery subsequent to the implantation of a bare-metal XIENCE-V stent. The initial phases of reendothelialization occur rather rapidly. However, the gradual and complete reestablishment of endothelial function takes a longer time, resulting in a substantial buildup of growth factors within the subintimal space. This, in turn, triggers an exacerbated growth response. The absence of the drug contributes to heightened restenotic growth as there are no anti-inflammatory effects within the arterial wall in this scenario although the endothelium recovers relatively quickly.

\begin{figure}[htbp!]
    \centering
    \includegraphics[scale=0.6]{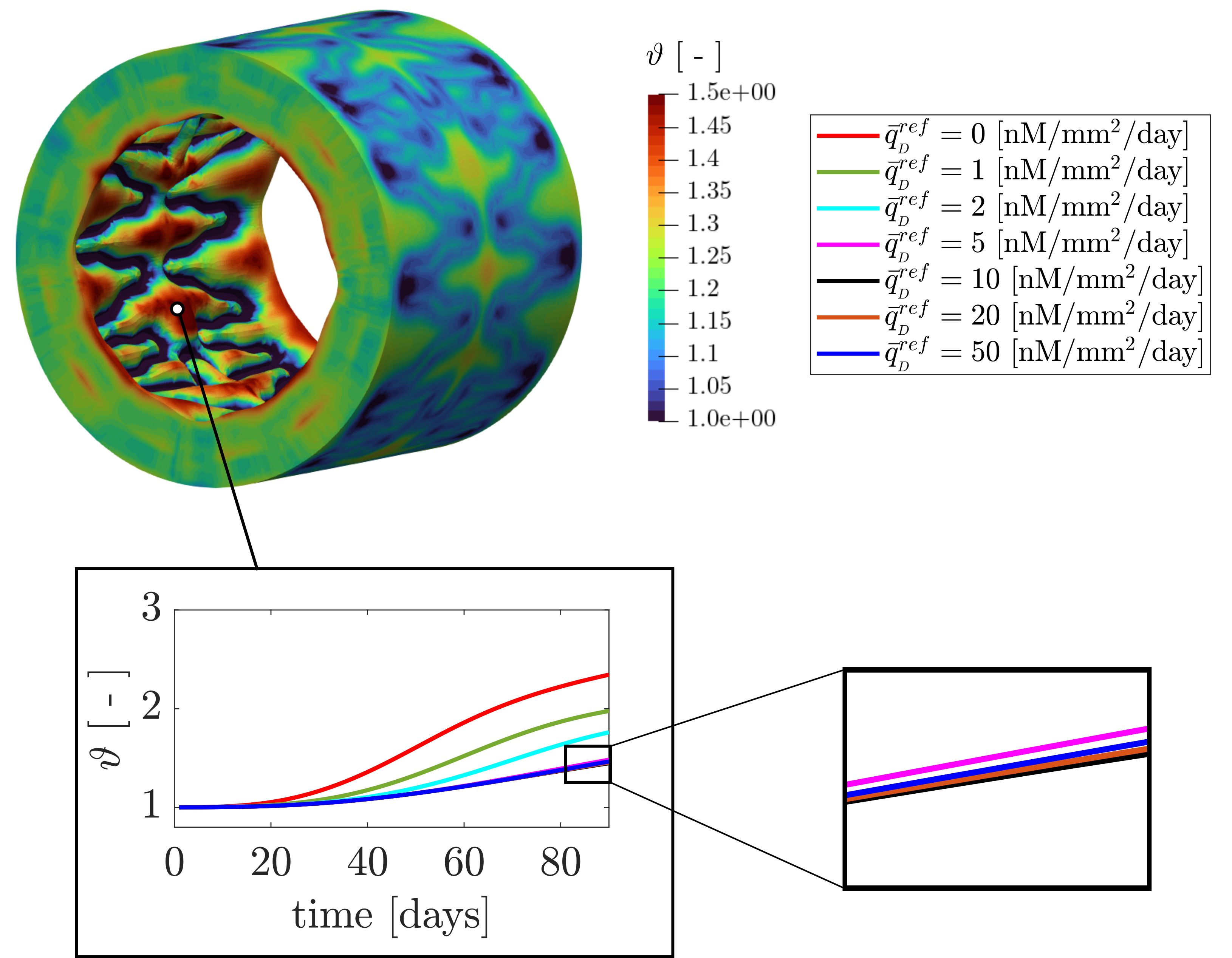}
    \caption{{Evolution of growth stretch $\vartheta$ [-] for varying levels of drug embedment.}}
    \label{fig_theta_evolution_IFAM}
\end{figure}

The impact of the drug level embedded in the stent struts, which is regulated by the parameter $\bar{q}^{ref}_{{}_D}$ representing the peak drug influx, is evident in Fig. \ref{fig_theta_evolution_IFAM}. This parameter significantly influences the extent of restenotic growth, as the presence of rapamycin leads to a delay in the healing of the endothelium. The results point towards an optimality in the aforementioned parameter.

\subsection{Hemodynamic effects}

In this section, we analyze the hemodynamics in stented arteries for a benchmark (ring stent) and a production case (ad-hoc XIENCE-V stent). We assume perfect contact between the stent and the artery and thus no indentation. All simulations are obtained with the highly parallelizable in-house code XNS \citep{Pauli2016b} and were performed on the supercomputers JURECA at Forschungszentrum Juelich \citep{krause2018jureca} and CLAIX 2018 at RWTH Aachen University.

\subsection{Idealized artery with a ring stent}
\label{ResultsRS}
We first compare the hemodynamics of a benchmark configuration, i.e., a ring stent with a square cross-section, to a healthy cylindrical artery segment. Both the stented and the healthy case have a radius of 1.8 mm and are 10 mm long. The ring stent thickness is 0.1 mm and we assume perfect contact between the stent and the artery. 

\begin{figure}[t]
    \centering
        \subfloat[Ring stent vortices.]{
        \resizebox*{!}{4.5cm}{
        \begin{tikzpicture}
            \node[anchor=south west,inner sep=0] (image) at (0,0) (image) {\includegraphics[draft=\draftmode]{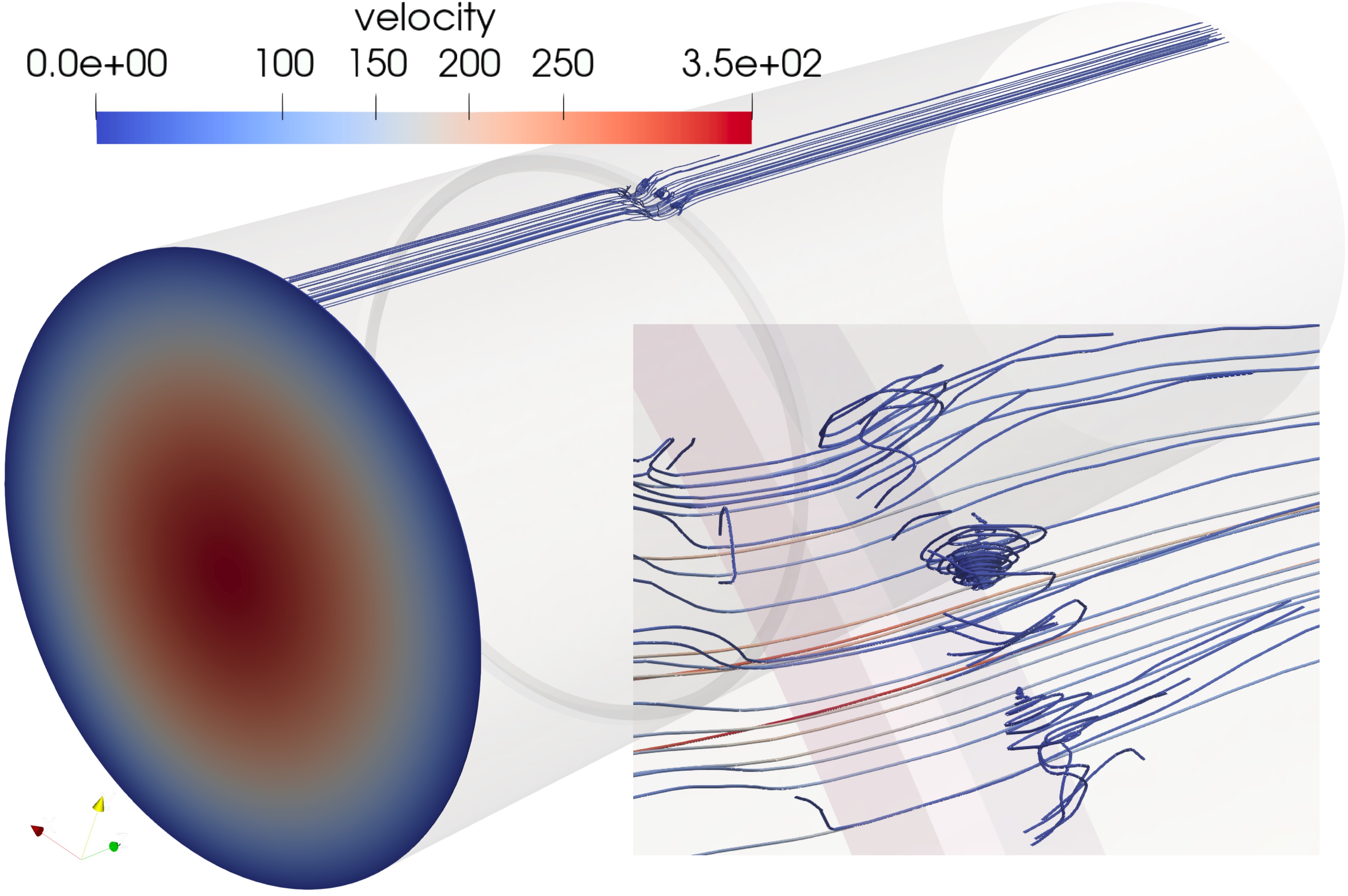}};
            \begin{scope}[x={(image.south east)},y={(image.north west)}]
                \draw[red, dashed, dash pattern = on 3cm off 3cm, line width=0.5cm] (0.47,0.04) rectangle (0.98,0.64);
                \draw[red, dashed, dash pattern = on 1cm off 1cm, line width=0.3cm] (0.46,0.75) rectangle (0.52,0.81);
            \end{scope}
        \end{tikzpicture}}
        \label{fig:streamlinesMain}}
        \subfloat[Velocity field.]{
        \resizebox*{6.3cm}{!}{
        \begin{tikzpicture}
            \node[anchor=south west,inner sep=0] (image) at (0,0) (image) {\includegraphics[draft=\draftmode]{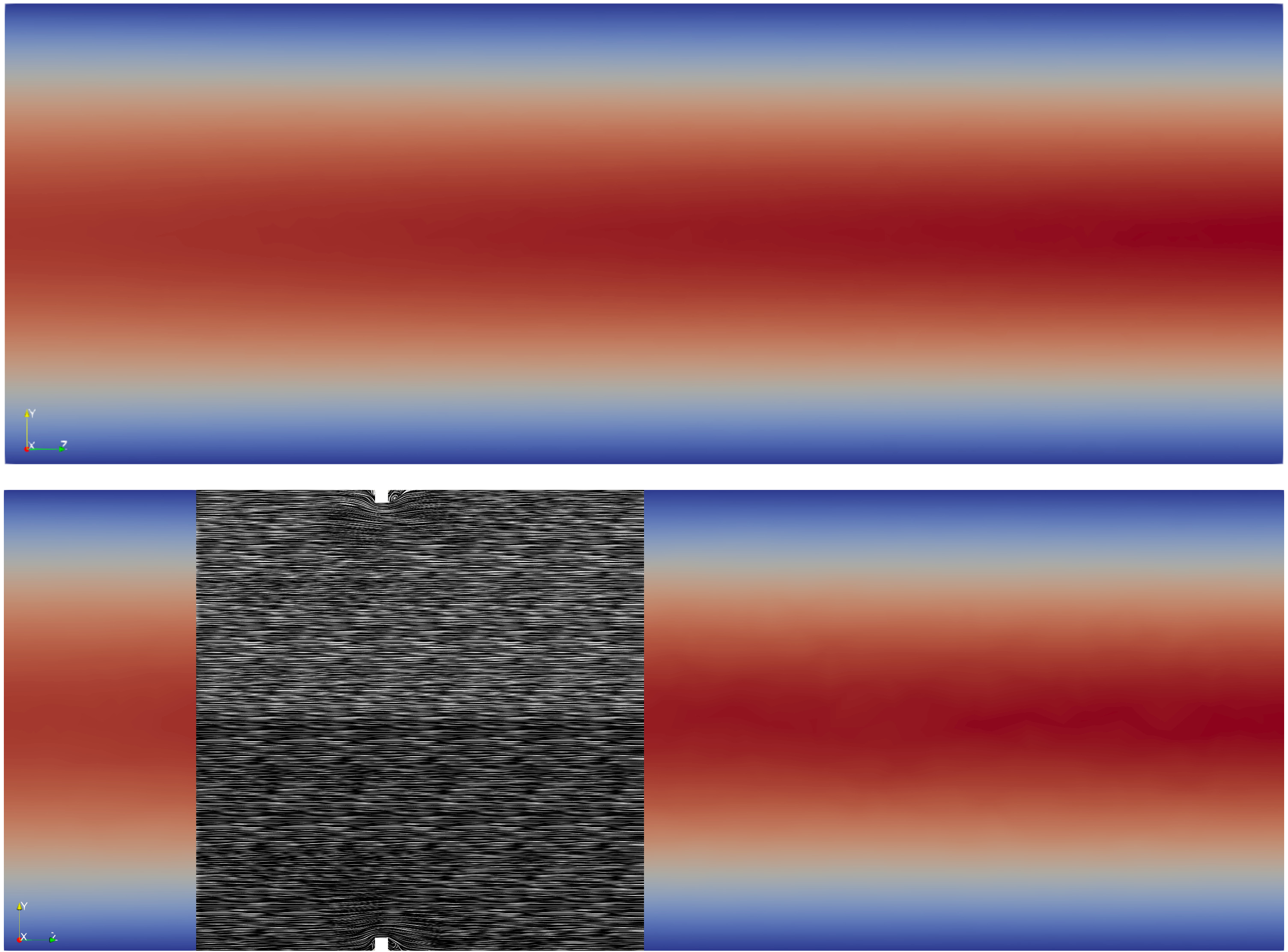}};
            \begin{scope}[x={(image.south east)},y={(image.north west)}]
                \draw[orange, line width=0.3cm] (0.28,0.454) rectangle (0.315,0.485);
                \draw[cyan, line width=0.3cm] (0.28,0) rectangle (0.315,0.029);
                \draw[green,line width=0.1cm,->] (0.31,0.5) -- (0.31,1);
                \draw[green, dashed, dash pattern = on 1cm off 1cm, line width=0.1cm,->] (0,0.75) -- (1,0.75);
            \end{scope}
        \end{tikzpicture}}
        \label{fig:surfaceLICMain}}
        \subfloat[Streamlines.]{
        \begin{tabular}[b]{c}
            \resizebox*{!}{2.25cm}{\begin{tikzpicture}
            \node[inner sep=0] (image) {\includegraphics[draft=\draftmode]{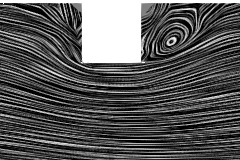}};
            \draw[orange, line width=0.2cm] (image.south west) rectangle (image.north east);
            \end{tikzpicture}}\\
            \resizebox*{!}{2.25cm}{\begin{tikzpicture}
            \node[inner sep=0] (image) {\includegraphics[draft=\draftmode]{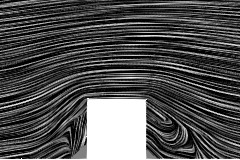}};
            \draw[cyan, line width=0.2cm] (image.south west) rectangle (image.north east);
            \end{tikzpicture}}
        \end{tabular}\label{fig:surfaceLICZoom}}\\
        \subfloat[Flow profiles over solid green line, see \ref{fig:streamlinesMain} (top).]
        {
        \resizebox*{!}{4.5cm}{
        \begin{tikzpicture}
            \node[anchor=south west,inner sep=0] (image) {\includegraphics[draft=\draftmode]{images/CATS/FlowProfilesComparison-new4}};
            \begin{scope}[x={(image.south east)},y={(image.north west)}]
            \draw[pink,line width=0.1cm] (0.15,0.17) circle (0.05 and 0.07);
            \draw[gray,line width=0.1cm] (0.94,0.17) circle (0.05 and 0.07);
            \end{scope}
        \end{tikzpicture}}
        \label{fig:flowProfilesMain}}
	   \subfloat[Zoom.]{
        \begin{tabular}[b]{c}
         \resizebox*{2.25cm}{!}{
         \begin{tikzpicture}
            \begin{scope}[x={(image.south east)},y={(image.north west)}]
            \clip (-0.35,-0.33) circle (0.05 and 0.07);
            \node[inner sep=0, anchor=center] at (0,0) {\includegraphics[draft=\draftmode]{images/CATS/FlowProfilesComparison-new4}};
            \draw[pink,line width=0.1cm] (-0.35,-0.33) circle (0.05 and 0.07);
            \end{scope}
        \end{tikzpicture}}\\
         \resizebox*{2.25cm}{!}{
         \begin{tikzpicture}
            \begin{scope}[x={(image.south east)},y={(image.north west)}]
            \clip (0.44,-0.33) circle (0.05 and 0.07);
            \node[inner sep=0, anchor=center] at (0,0) {\includegraphics[draft=\draftmode]{images/CATS/FlowProfilesComparison-new4}};
            \draw[gray,line width=0.1cm] (0.44,-0.33) circle (0.05 and 0.07);
            \end{scope}
        \end{tikzpicture}}
        \end{tabular}
	   \label{fig:flowProfilesZoom}}
    \subfloat[Rectangular stent]{
        \begin{tabular}[b]{c}
         \resizebox*{!}{1.5cm}{
         \includegraphics[draft=\draftmode]{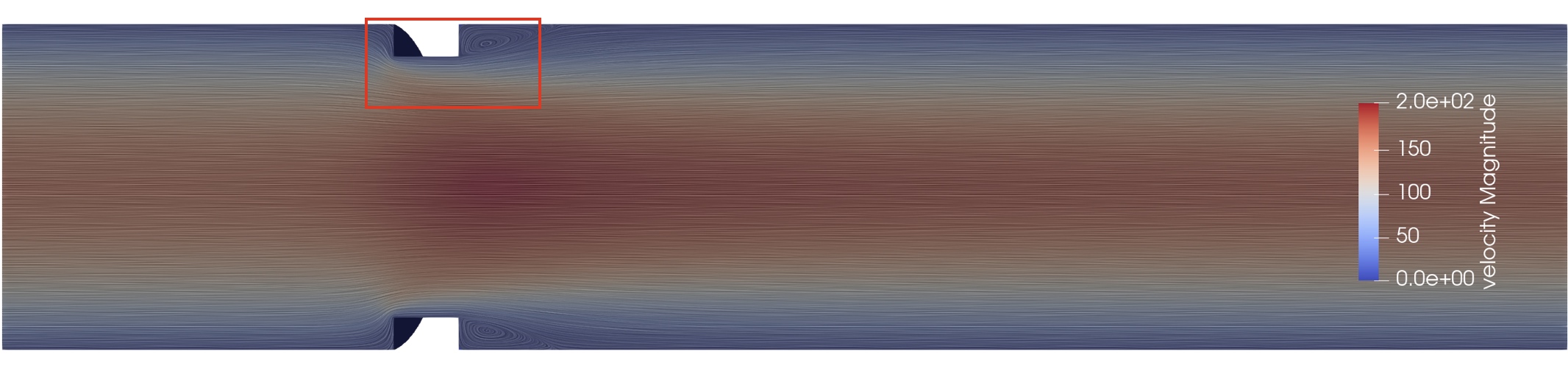}}\\
         \resizebox*{!}{3cm}{
         \includegraphics[draft=\draftmode]{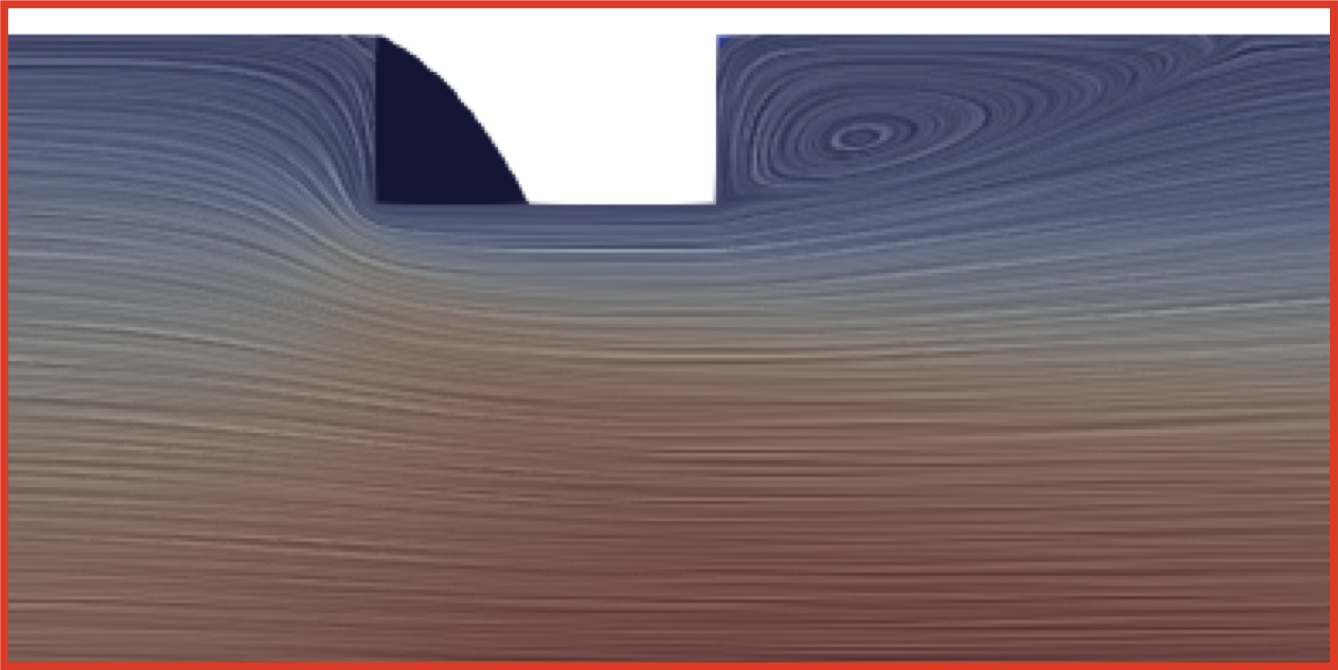}}
        \end{tabular}
	   \label{fig:ringComparisonZoom}}
    \caption{Velocity field comparison between stented and healthy artery: local vortices at t = 0.25 s with a ring stent in \ref{fig:streamlinesMain}; velocity magnitude and streamlines at t = 0.25 s in \ref{fig:surfaceLICMain}, solid green line for flow profile comparison in \ref{fig:flowProfilesMain} and centerline marked with a dashed green line; corresponding zoom on streamlines for ring stent in \ref{fig:surfaceLICZoom}; comparison of flow profiles at different time snapshots, flow in the healthy case is marked with dashed lines and in the stented case (RS) with solid lines (\ref{fig:flowProfilesMain} and \ref{fig:flowProfilesZoom}). Streamlines and local vortex comparison with rectangular ring stent in \ref{fig:ringComparisonZoom}. }
    \label{fig:surfaceLIC}
\end{figure}

Different indentation levels for this test case have been investigated in \citep{Nerzak}. Figure \ref{fig:surfaceLICMain} shows the velocity field at time $t=0.25$ s in the artery segment with and without stent. If we compare velocity profiles at different time steps in Fig. \ref{fig:flowProfilesMain} and \ref{fig:flowProfilesZoom}, we observe that the flow is affected only in the stent proximity. In particular, bigger vortices appear downstream of the ring stent, see Fig. \ref{fig:streamlinesMain} and \ref{fig:surfaceLICZoom} and their size is strictly dependent on the size of the strut. Figures \ref{fig:ringComparisonZoom} highlight the different vortex scales with a larger, rectangular stent. The ring stent can be regarded as a single strut in a complex stent design. Thus, we expect to observe similar hemodynamics close to all struts of the XIENCE-V stent.\\
\newsavebox{\myimagebox}
\savebox{\myimagebox}{\includegraphics{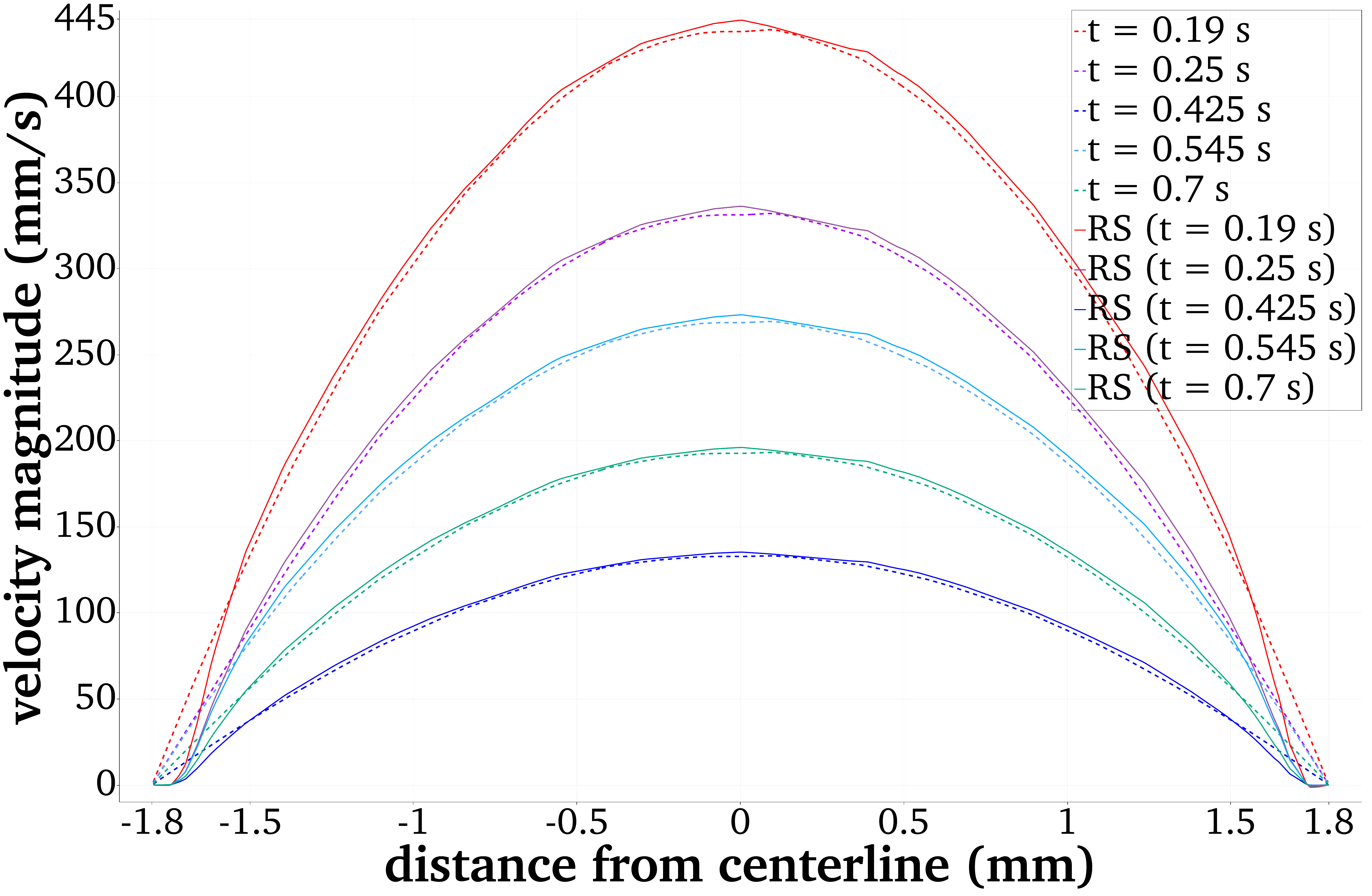}}
\newlength{\myimagewidth}
\newlength{\myimageheight}
\setlength{\myimagewidth}{\wd\myimagebox}
\setlength{\myimageheight}{\ht\myimagebox}

An important parameter to analyze the well-being of the artery wall is the WSS: healthy values of WSS are between 0.8 and 1.5 Pa \citep{kim2011shear,maul2011mechanical}. Thus, we investigate the areas of low WSS where blood recirculation and induced inflammation is expected. The WSS measured on the healthy case in Fig. \ref{fig:WSSPipe} corresponds to the physiological baseline. Figures \ref{fig:WSSRS} and \ref{fig:WSSStentedWall} display $\wss_1$ on the artery wall and on the stent. In particular, we observe that the stent is subject to very high WSS while the wall immediately downstream of the stent has critically low WSS, in agreement with the results in \citep{koskinas2012role}. By definition, $\wss_1$ and $\wss_2$ are equivalent in the healthy case but can show significant differences in the presence of a ring stent. In figure \ref{fig:ErrorWSSRS} we show the relative error
\begin{equation}
    \epsilon_{\wss} = \frac{|\wss_1-\wss_2|}{\wss_2}.
\end{equation}
\definecolor{myyellow}{RGB}{255,170,0}
\definecolor{mygreen}{RGB}{0,170,127}
\begin{figure} [t]
    \centering
    \subfloat[WSS on ring stent.]{
    \resizebox*{!}{3.5cm}{
    \begin{tikzpicture}
        \node[anchor=south west,inner sep=0] (image) at (0,0) (image) {\includegraphics[draft=\draftmode]{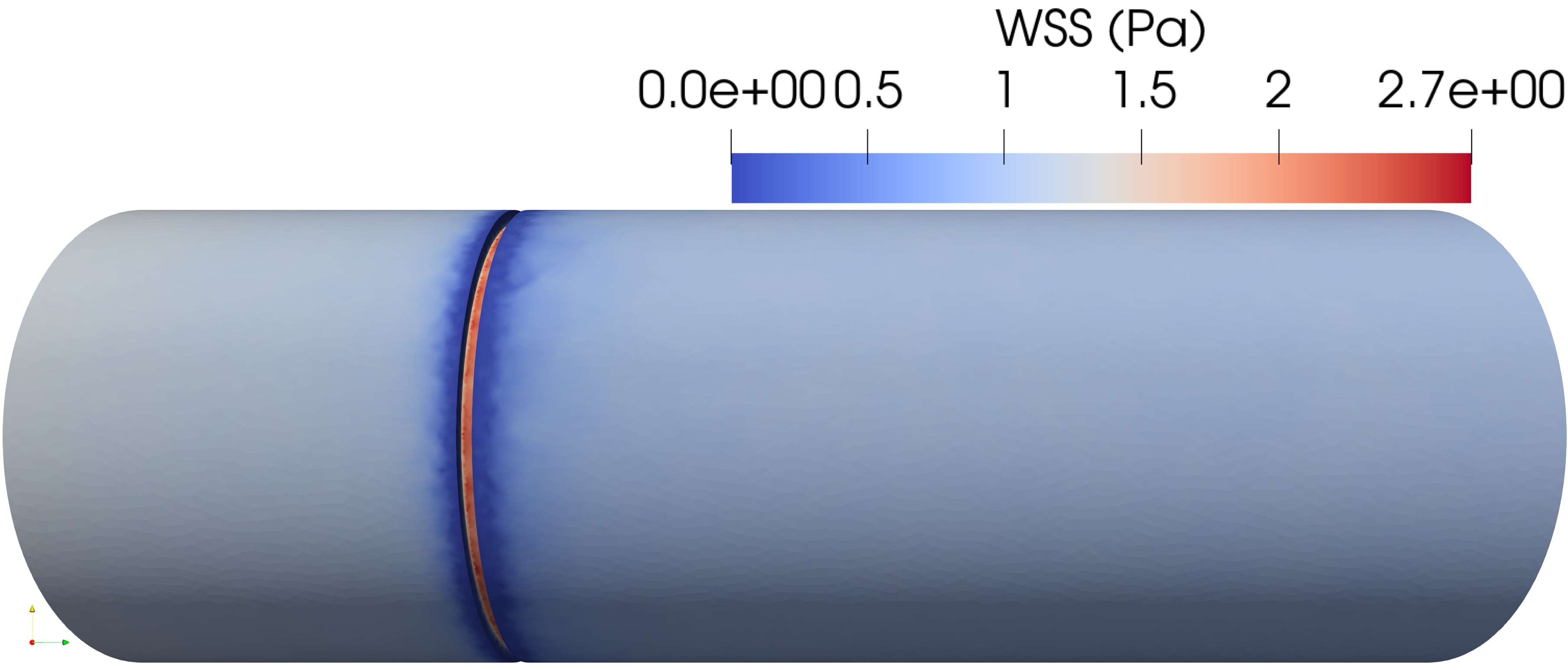}};
        \begin{scope}[x={(image.south east)},y={(image.north west)}]
            \draw[black,line width=0.1cm, fill=mygreen] (0.03,0.35) circle (0.009 and 0.02);
            \draw[black,line width=0.1cm, fill=blue] (0.95,0.35) circle (0.009 and 0.02);
            \draw[black,line width=0.1cm, fill=myyellow] (0.27,0.35) circle (0.009 and 0.02);
            \draw[black,line width=0.1cm, fill=red] (0.33,0.35) circle (0.009 and 0.02);
        \end{scope}
    \end{tikzpicture}}
    \label{fig:WSSRS}}
    \hspace{3pt}
    \subfloat[WSS on healthy artery.]{
    \resizebox*{!}{3.5cm}{
    \begin{tikzpicture}
        \node[anchor=south west,inner sep=0] (image) at (0,0) (image) {\includegraphics[draft=\draftmode]{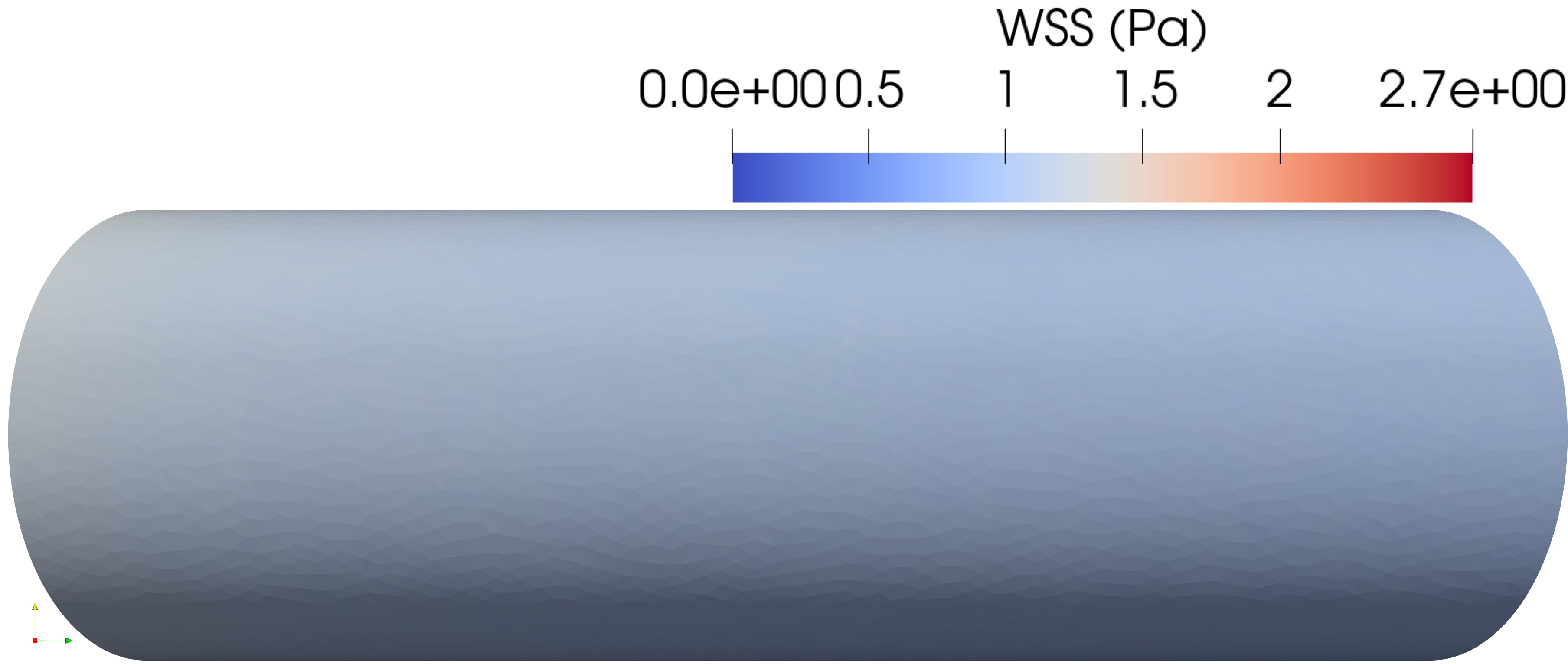}};
        \begin{scope}[x={(image.south east)},y={(image.north west)}]
            \draw[black,line width=0.1cm, fill=mygreen] (0.03,0.35) circle (0.009 and 0.02);
            \draw[black,line width=0.1cm, fill=blue] (0.95,0.35) circle (0.009 and 0.02);
            \draw[black,line width=0.1cm, fill=myyellow] (0.3,0.35) circle (0.009 and 0.02);
        \end{scope}
    \end{tikzpicture}}
    \label{fig:WSSPipe}}
    \hfill
    \subfloat[WSS on stented wall.]{
    \resizebox*{!}{3.5cm}{
    \begin{tikzpicture}
        \node[anchor=south west,inner sep=0] (image) at (0,0) (image) {\includegraphics[draft=\draftmode]{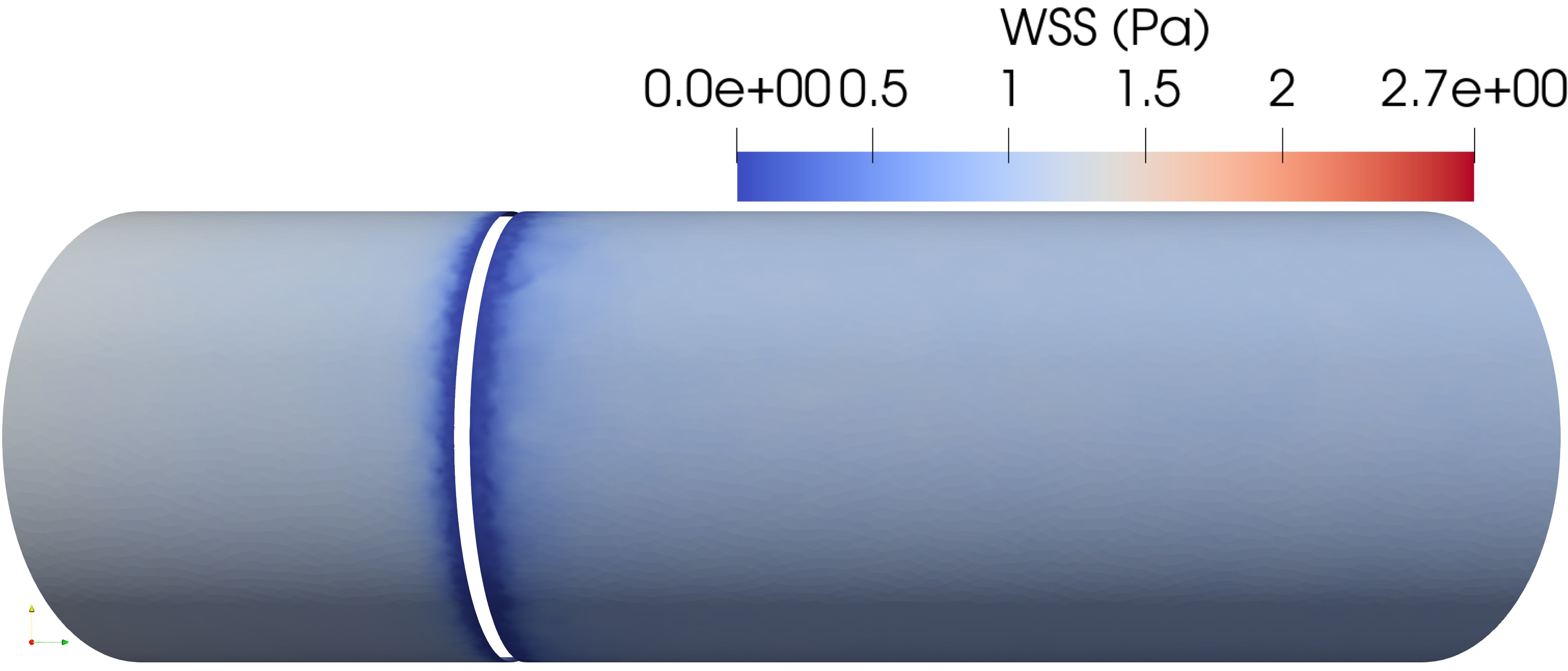}};
        \begin{scope}[x={(image.south east)},y={(image.north west)}]
            \draw[black, line width=0.5cm] (0.25,0) rectangle (0.35,0.7);
        \end{scope}
    \end{tikzpicture}}
    \label{fig:WSSStentedWall}}
    \hspace{3pt}
    \subfloat[$\epsilon_{\wss}$ on stented wall.]{
    \resizebox*{!}{3.5cm}{\includegraphics[draft=\draftmode]{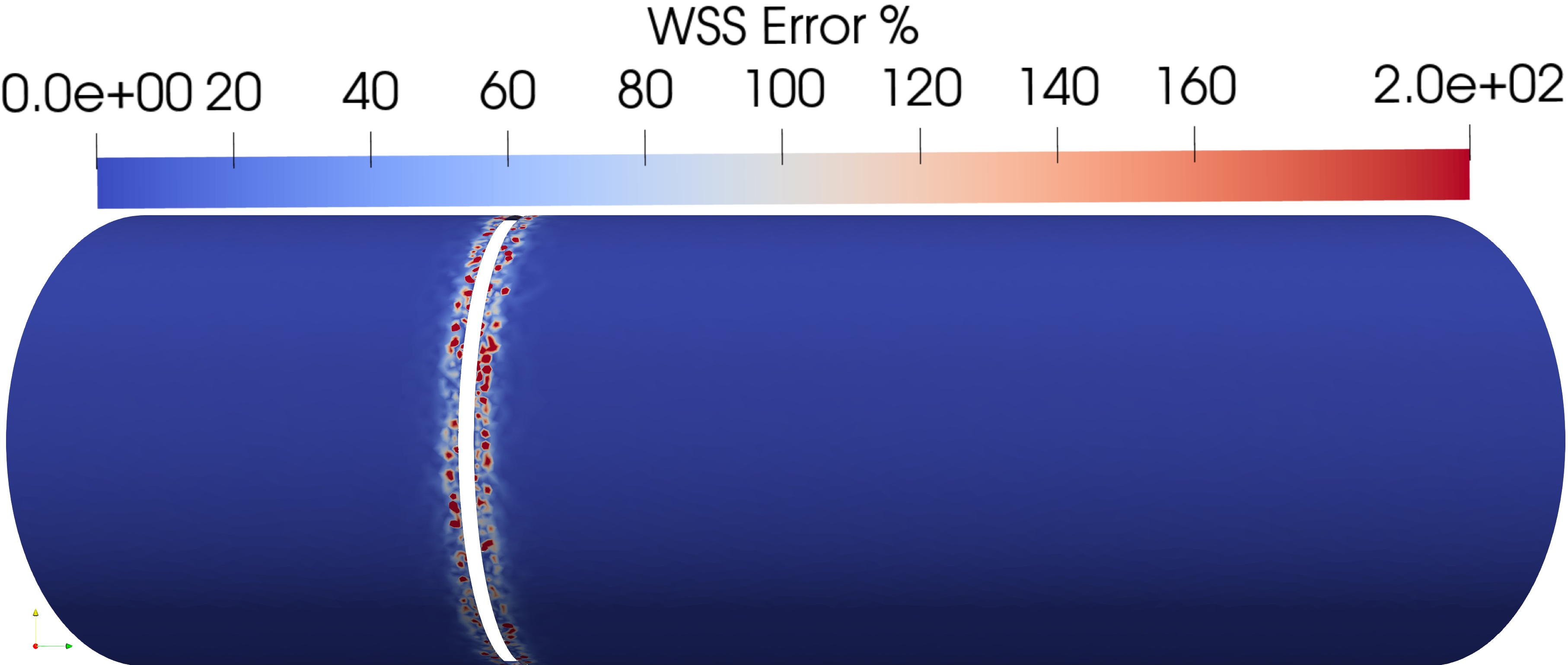}}
    \label{fig:ErrorWSSRS}}
    \caption{WSS and $\epsilon_{\wss}$ on stented and healthy artery. The colored dots in \ref{fig:WSSRS} and \ref{fig:WSSPipe} correspond to the plot colors in figure \ref{fig:WSSComparison}. From \ref{fig:WSSStentedWall}, zoom on rectangle can be found in figure \ref{fig:ZoomRS}.}
    \label{fig:WSSPipeRS}
\end{figure}
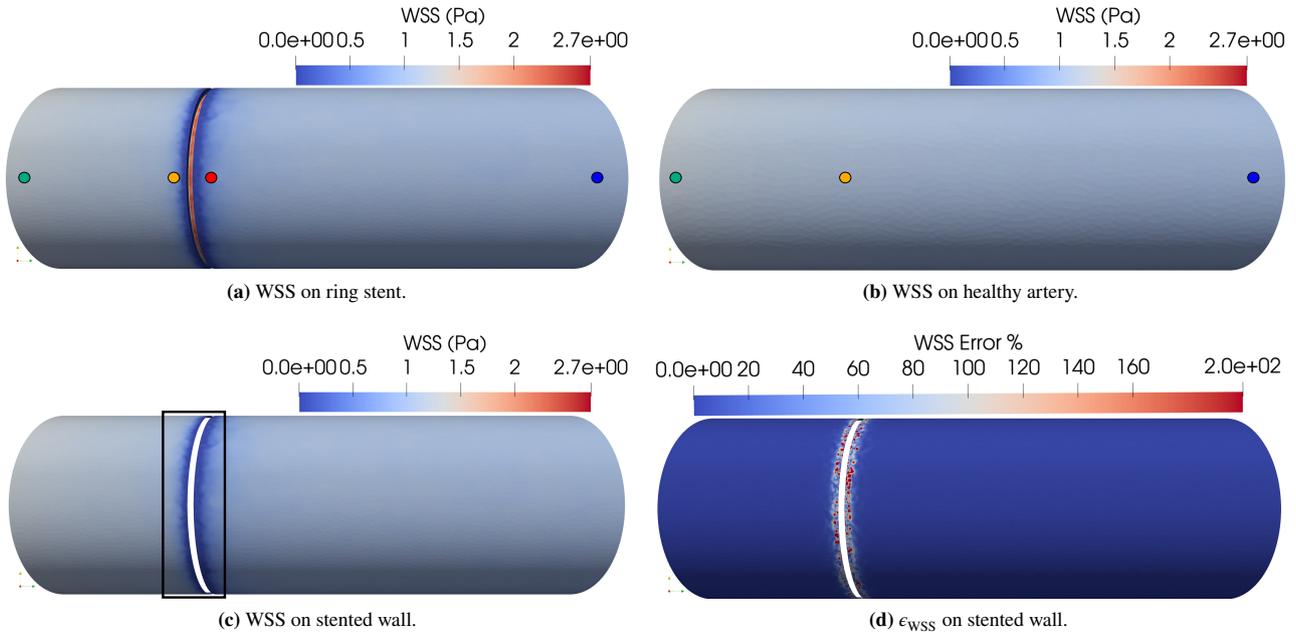

\begin{figure}[htbp!]
    \centering
    \subfloat[WSS for healthy and ring stent (RS) case.]{
    \resizebox*{!}{4cm}{\includegraphics[draft=\draftmode]{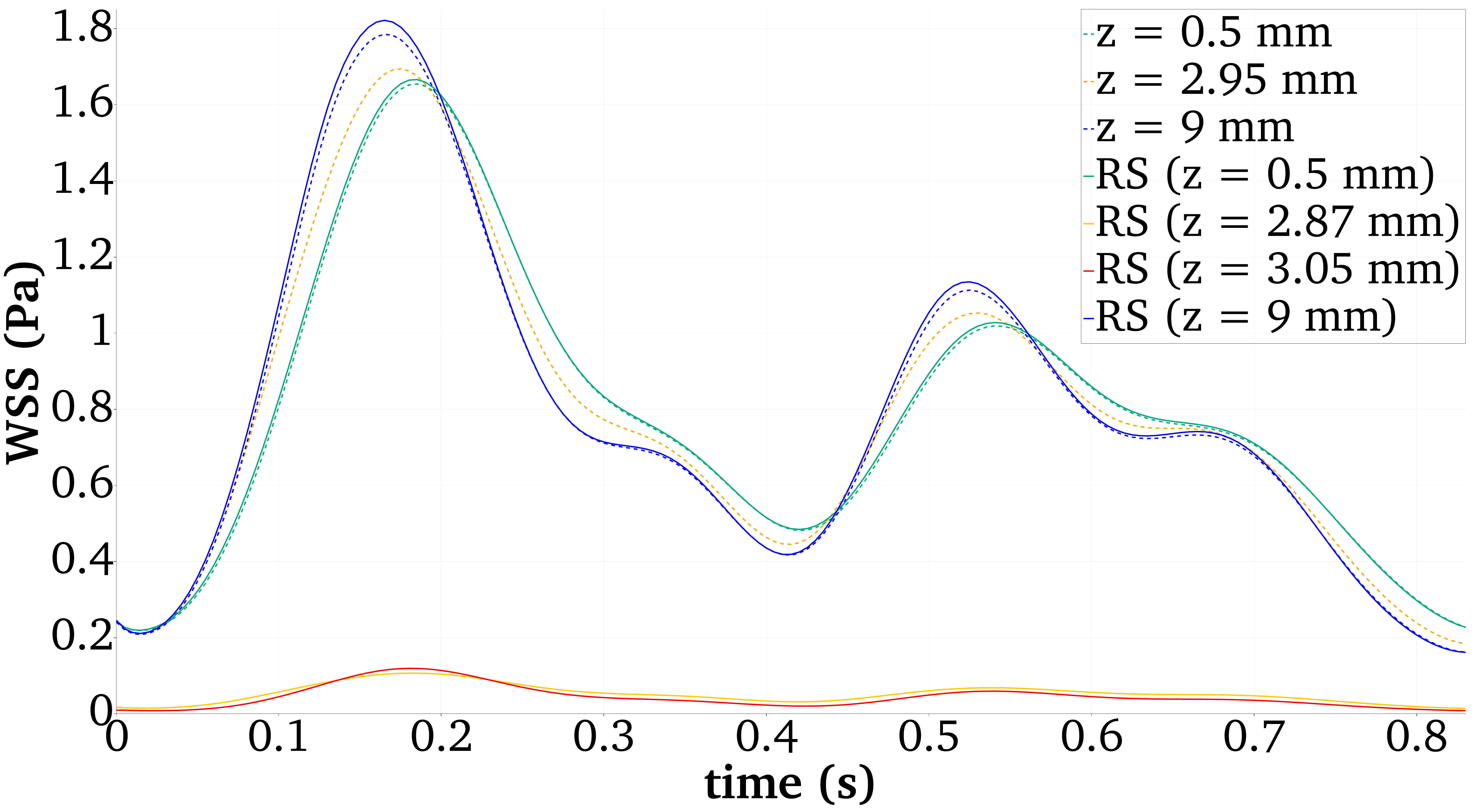}}
    \label{fig:WSSComp}}
    \hspace{3pt}
    \subfloat[$\wss_1$, $\wss_2$ comparison for RS.]{
    \resizebox*{!}{4cm}{\includegraphics[draft=\draftmode]{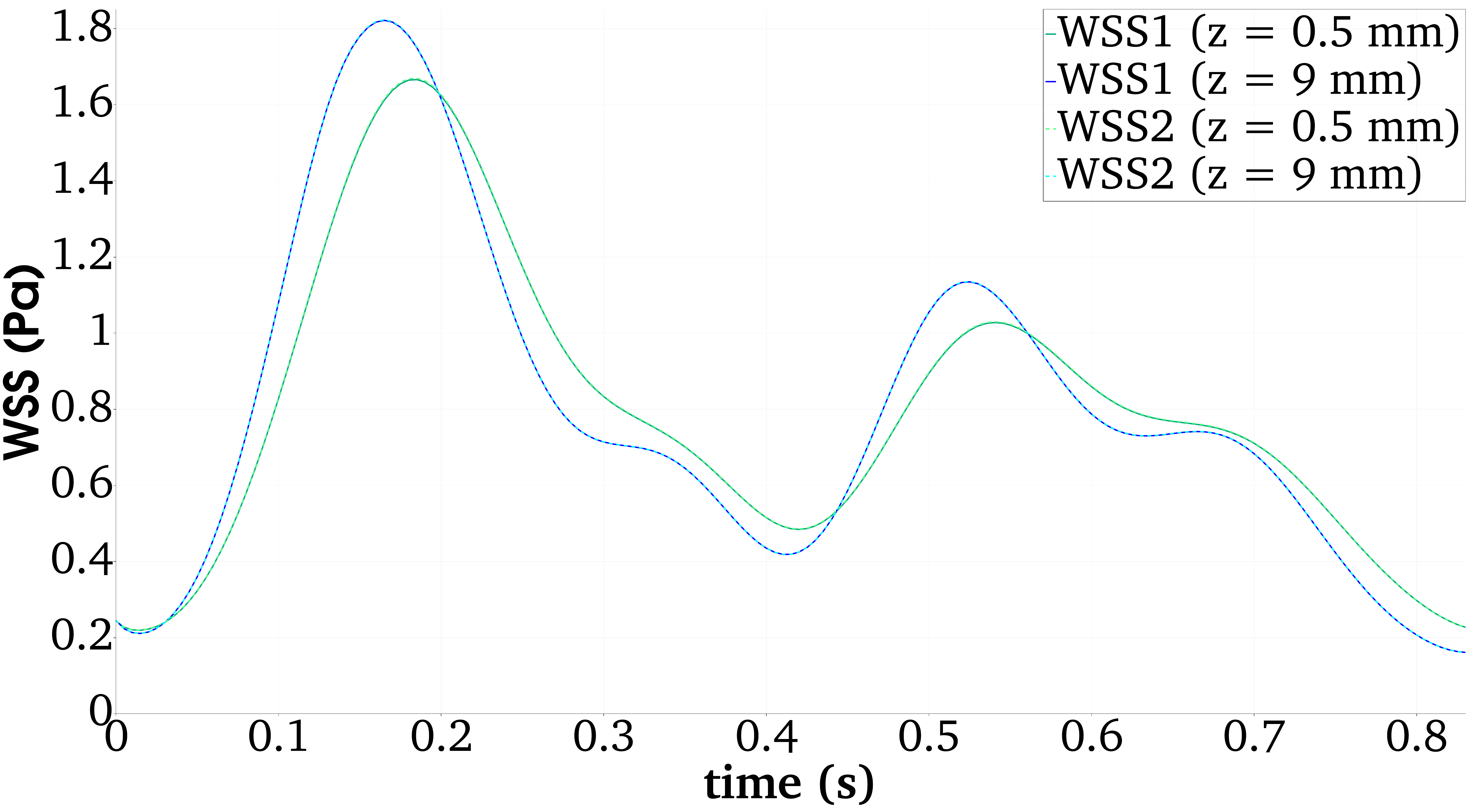}}
    \label{fig:WSSComp2}}\\
    \subfloat[$\wss_1$, $\wss_2$ comparison for RS in stent vicinity.]{
    \resizebox*{!}{4cm}{\includegraphics[draft=\draftmode]{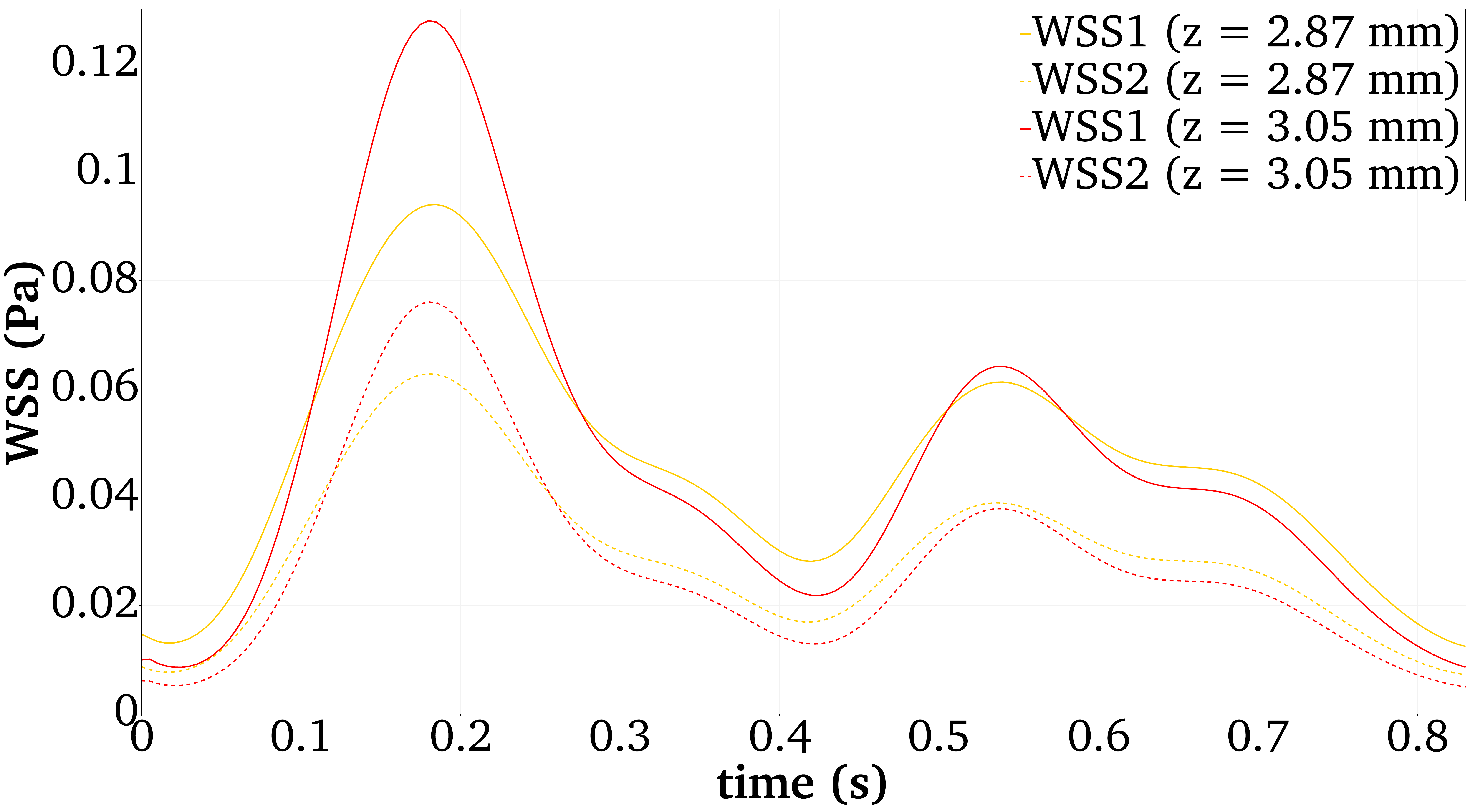}}
    \label{fig:WSSComp3}}
    \hspace{3pt}
    \subfloat[$\epsilon_{\wss}$ for RS in the stent vicinity.]{
    \resizebox*{!}{4cm}{\includegraphics[draft=\draftmode]{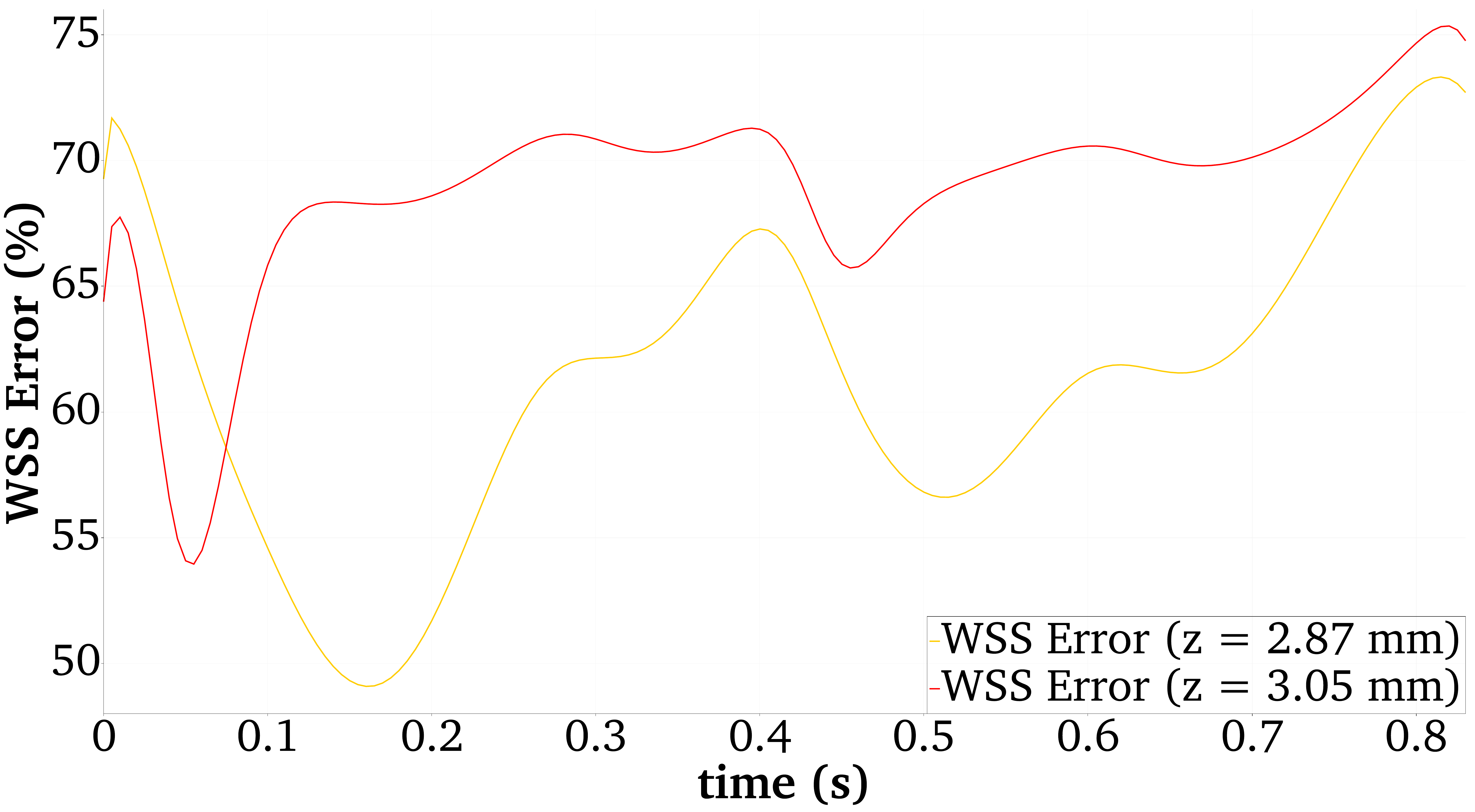}}
    \label{fig:WSSError}}
    \caption{WSS and error comparison upstream, downstream and in the stent vicinity over one cycle T. Lines are color-coded according to the location in Fig. \ref{fig:WSSPipeRS}.}
    \label{fig:WSSComparison}
\end{figure}

We pick four points on the artery wall to compare against the physiological values and to assess the influence of $\wss_1$ and $\wss_2$ over one heart beat. Figures \ref{fig:WSSComp} and \ref{fig:WSSComp2} show that the stented artery has physiological values of WSS near inflow and outflow and that the computation of $\wss_*$ is negligible far away from the stent. However, WSS values are critically low in the stent proximity and $\wss_1$ highly overestimates WSS values in the strut vicinity, see Fig. \ref{fig:WSSComp3}. The relative error in Fig. \ref{fig:WSSError} is between 40\% and 70\% over one heart beat and Fig. \ref{fig:ErrorWSSRS} shows peaks of 200\% in other nodes close to the ring stent. The error distribution is scattered due to the fact that many nodes adjacent to the ring stent have zero values of WSS and therefore $\epsilon_{\wss}$ was set to zero.\\

\begin{figure}[htbp!]
    \centering
    \subfloat[TAWSS.]{
    \resizebox*{4.5cm}{!}{\includegraphics[draft=\draftmode]{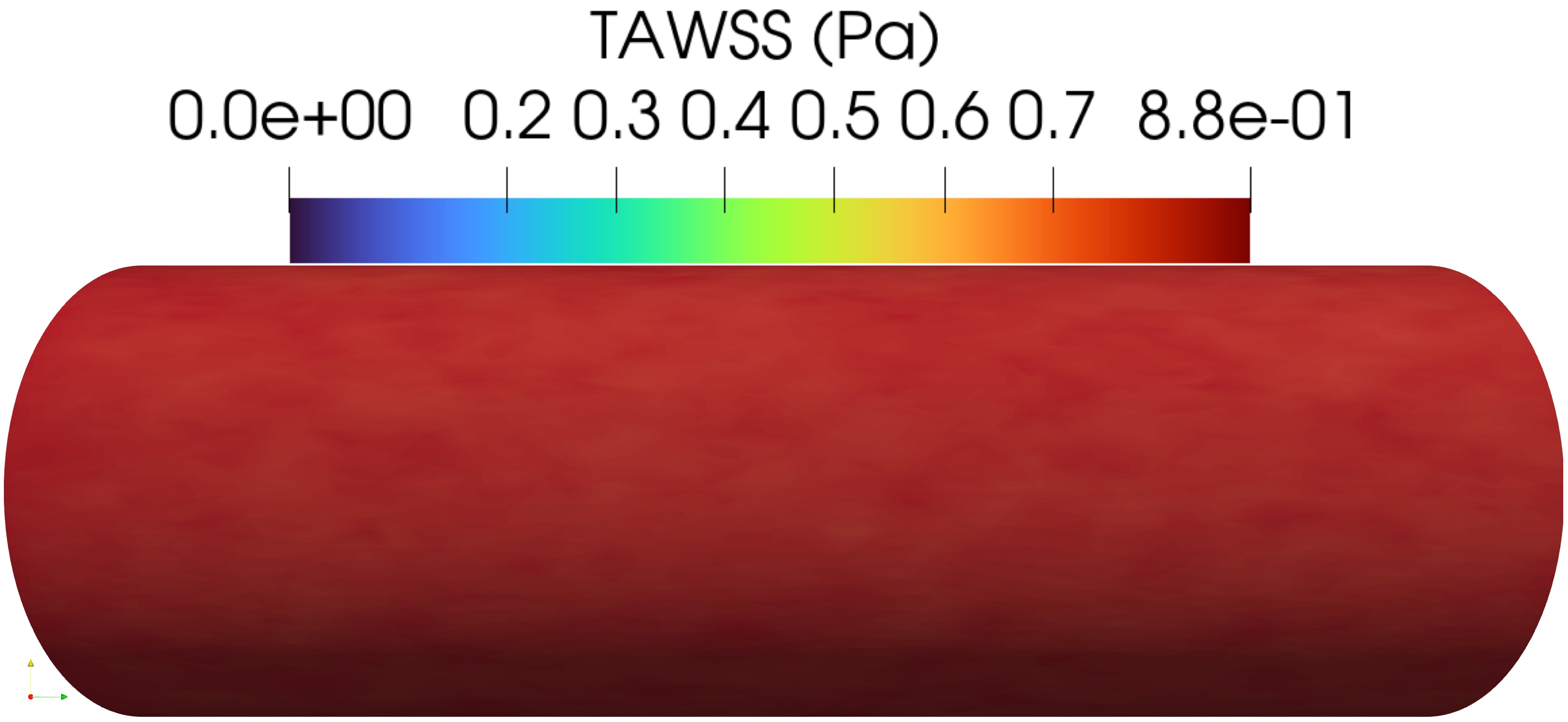}}
    }
    \hfill
    \subfloat[OSI.]{
    \resizebox*{4.5cm}{!}{\includegraphics[draft=\draftmode]{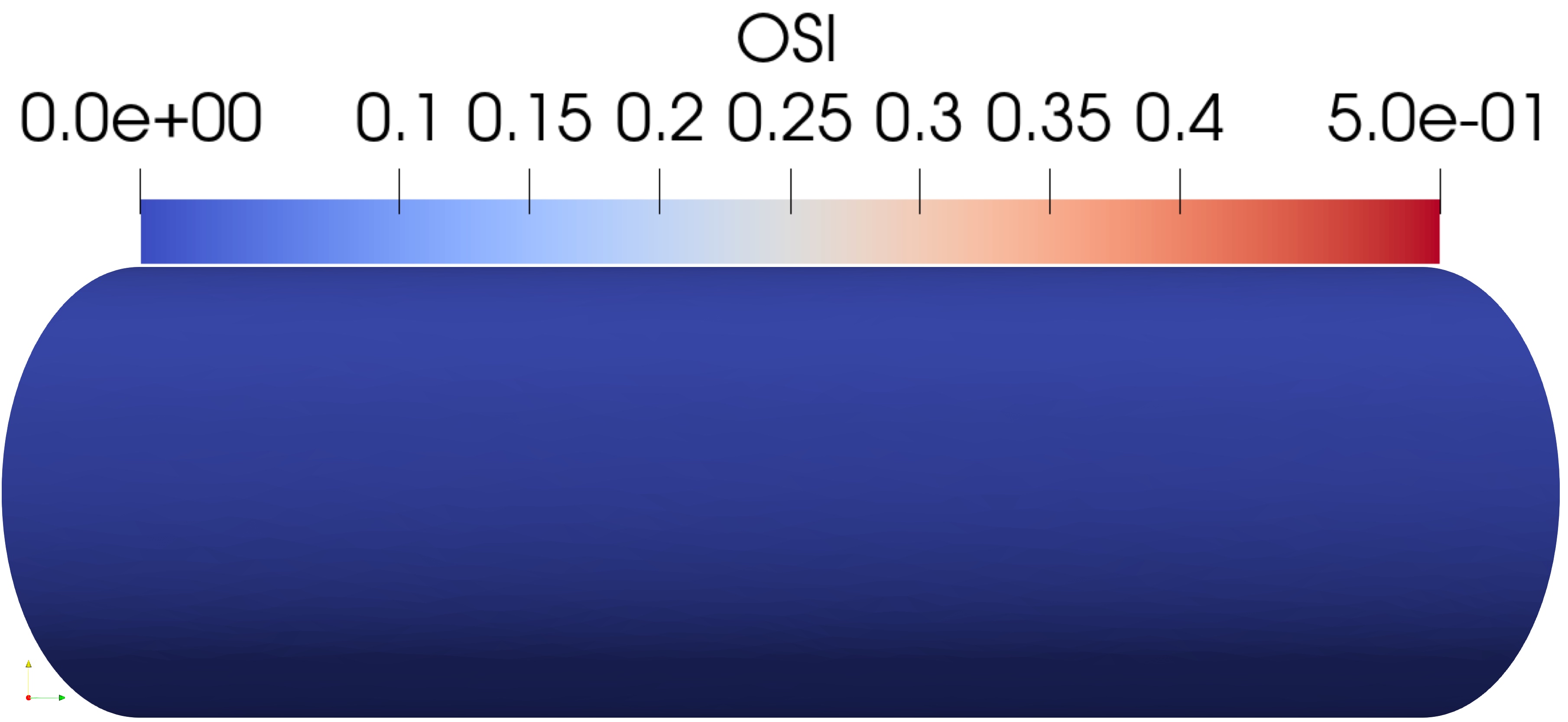}}
    }
    \hfill
    \subfloat[RRT.]{
    \resizebox*{4.5cm}{!}{\includegraphics[draft=\draftmode]{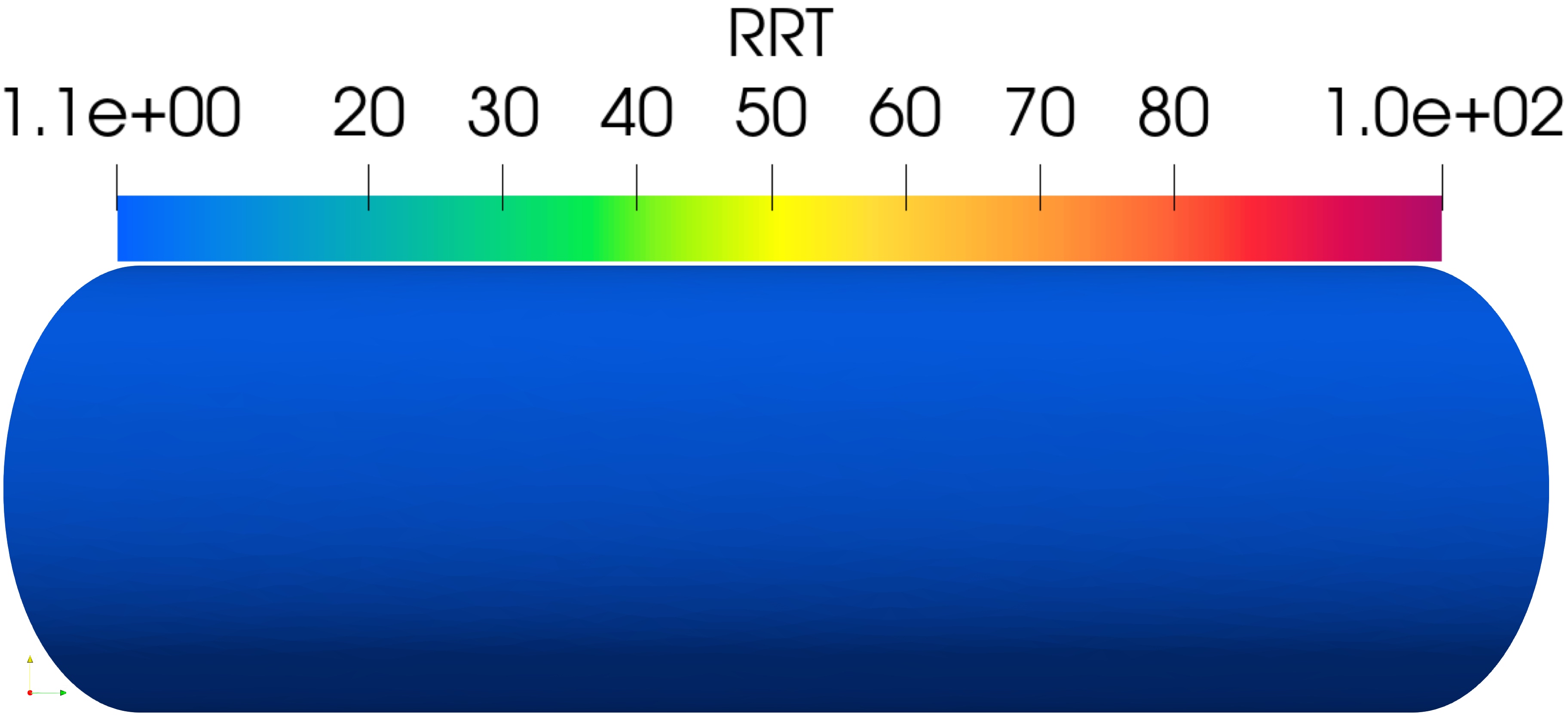}}
    }
    \caption{Physiological values on healthy artery.}
    \label{fig:physioSol}
\end{figure}

We also tested the influence of the relative error on time-integrated quantities such as time-averaged WSS (TAWSS), oscillatory shear index (OSI), and relative residence time (RRT) given by the following expressions:
\begin{equation}
	\tawss= \frac{1}{T}\int_{0}^{T} |\bm{\tau} (t)|dt,
	\label{TAWSS}
\end{equation}
\begin{equation}
	\osi = \frac{1}{2}\left(1-\frac{|\int_{0}^{T} \bm{\tau} (t)dt|}{\int_{0}^{T} |\bm{\tau} (t)|dt}\right),
	\label{OSI}
\end{equation}
\begin{equation}
\rrt = \frac{1}{(1-2\hspace{2mm}\osi)\tawss},
\label{RRT}
\end{equation}
where $\bm{\tau} = \bm{\tau}_*$ and $*=1,2$.
Figure \ref{fig:physioSol} shows physiological values of approximately 0.8 Pa TAWSS, zero OSI, and RRT between $1.2$ and $1.3$ $\frac{1}{Pa}$.

\begin{figure}[htbp!]
    \centering
    \subfloat[Error $\epsilon_{\tawss}$.]{
    \resizebox*{4.5cm}{!}{\includegraphics[draft=\draftmode]{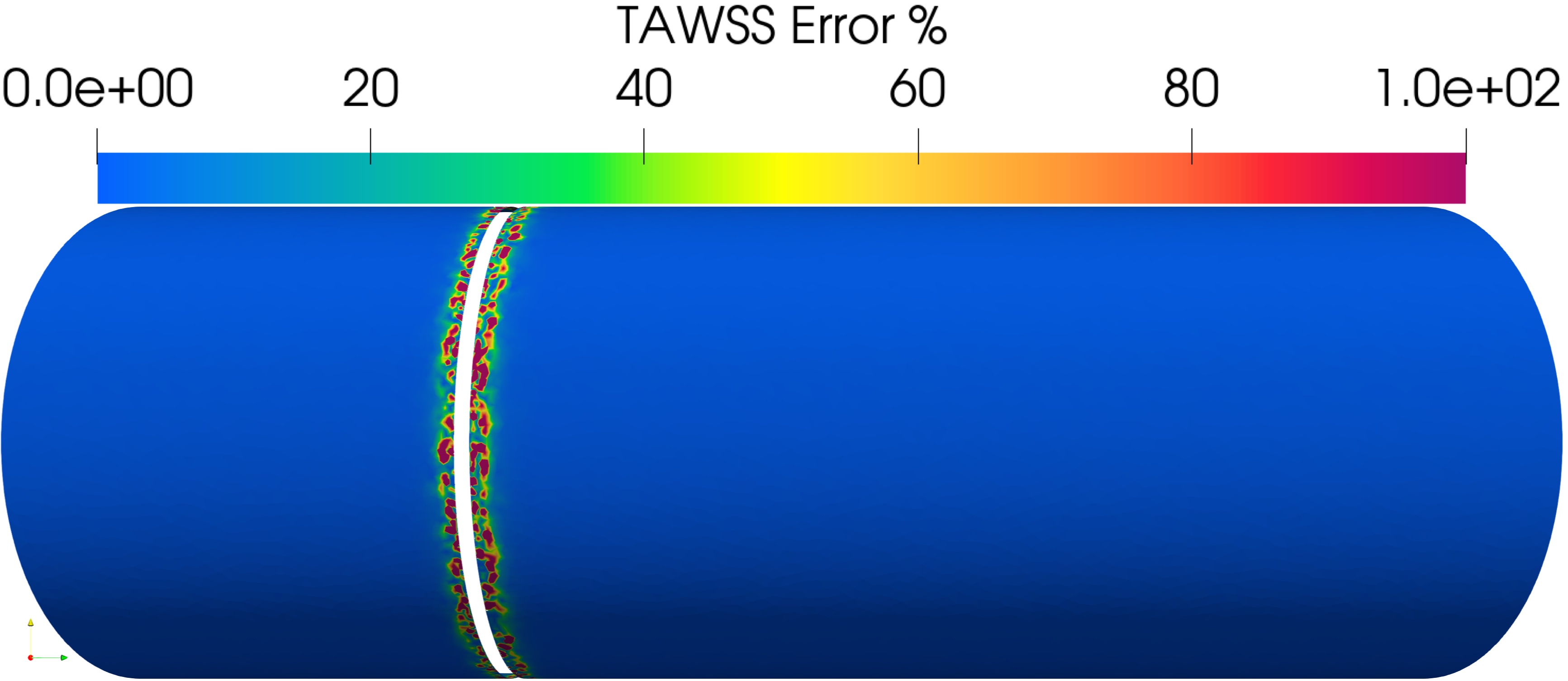}}
    }
    \hfill
    \subfloat[Deviation $\delta_1$.]{
    \resizebox*{4.5cm}{!}{\includegraphics[draft=\draftmode]{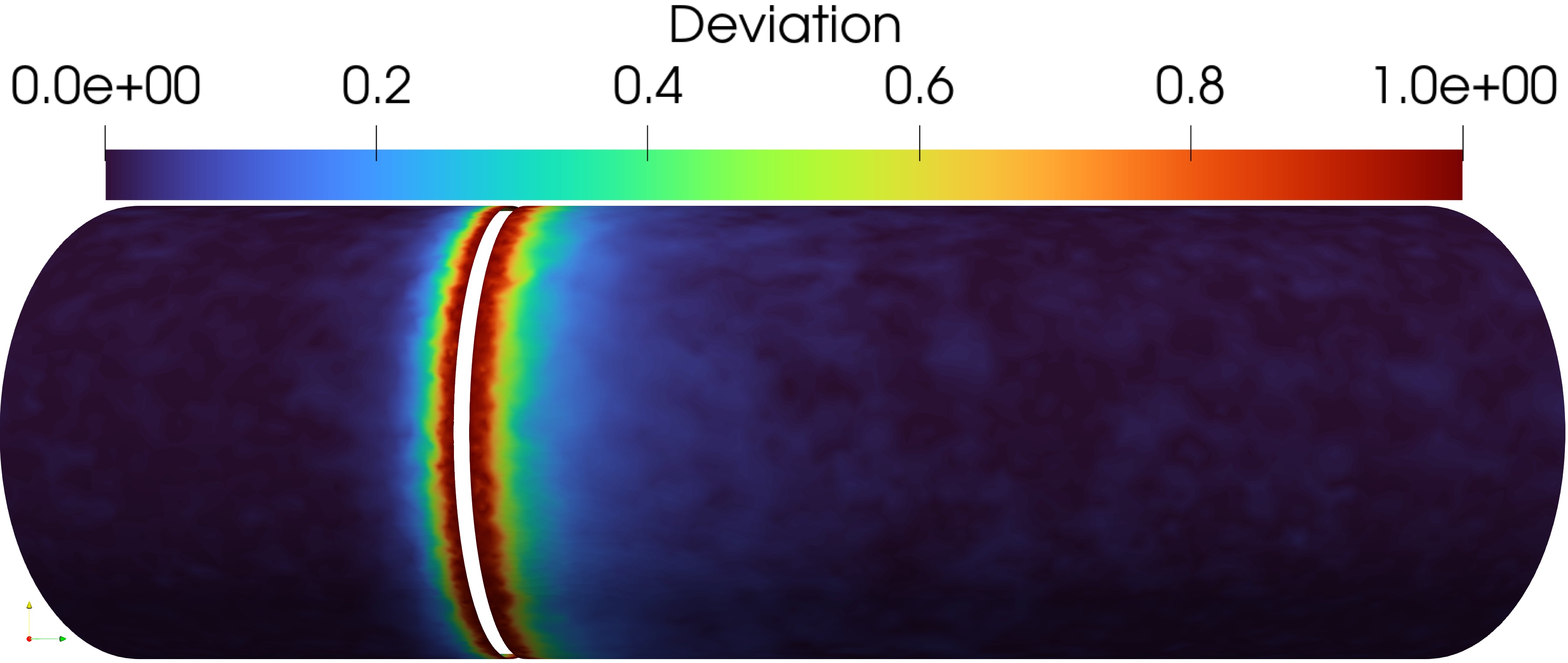}}
    }
    \hfill
    \subfloat[Deviation $\delta_2$.]{
    \resizebox*{4.5cm}{!}{\includegraphics[draft=\draftmode]{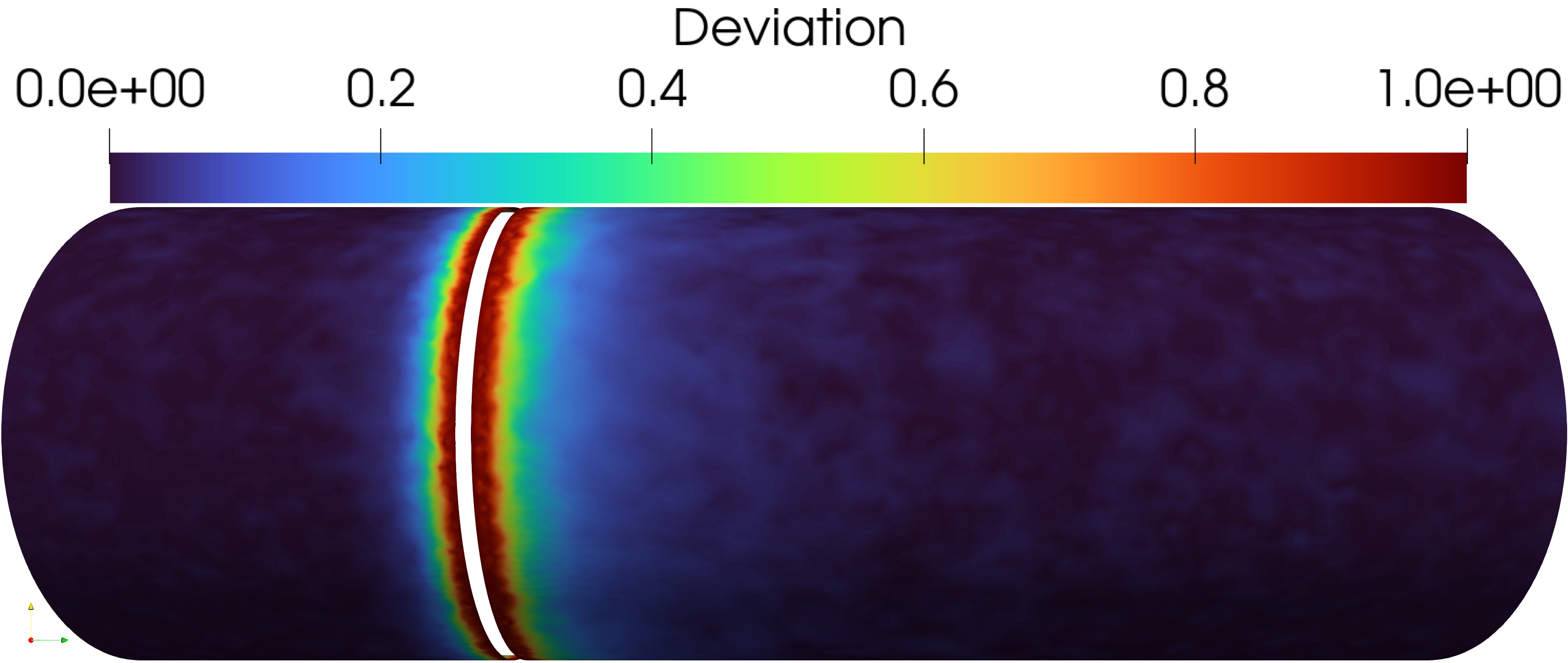}}
    }
    \caption{TAWSS error and deviation in artery with ring stent.}
    \label{fig:TAWSSErrorDevRS}
\end{figure}

These indicators are also compared on the artery wall \review{for both $\wss_*$ computations.} It is worth pointing out that although the relative error can be locally very high, the normalized relative deviation
\begin{equation}
    \delta_* = \frac{|\tawss_* - \tawss_{phys}|}{\tawss_{phys}}
\end{equation}
where $\tawss_{phys}=0.8$ Pa, is very similar for both computations of TAWSS \review{(see Figure \ref{fig:TAWSSErrorDevRS})}. This entails that the highest errors correspond to very low values of WSS ($10^{-3}$ to $10^{-5}$) and that $\wss_1$ overestimates shear stresses by one or two orders of magnitude.

\begin{figure}[htbp!]
    \centering
    \subfloat[$\tawss_1$]{
    \resizebox*{4.5cm}{!}{\includegraphics[draft=\draftmode]{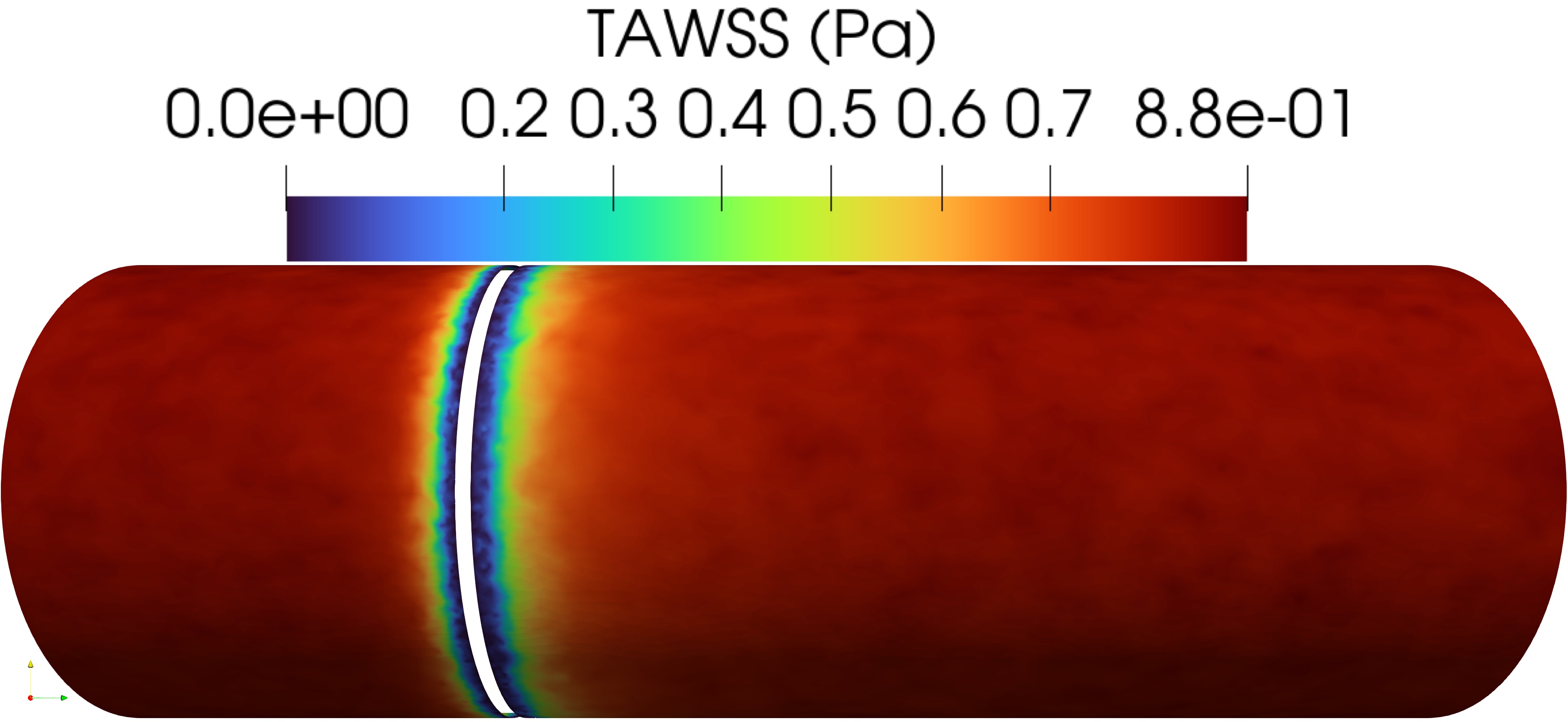}}
    }
    \hfill
    \subfloat[$\osi_1$]{
    \resizebox*{4.5cm}{!}{\includegraphics[draft=\draftmode]{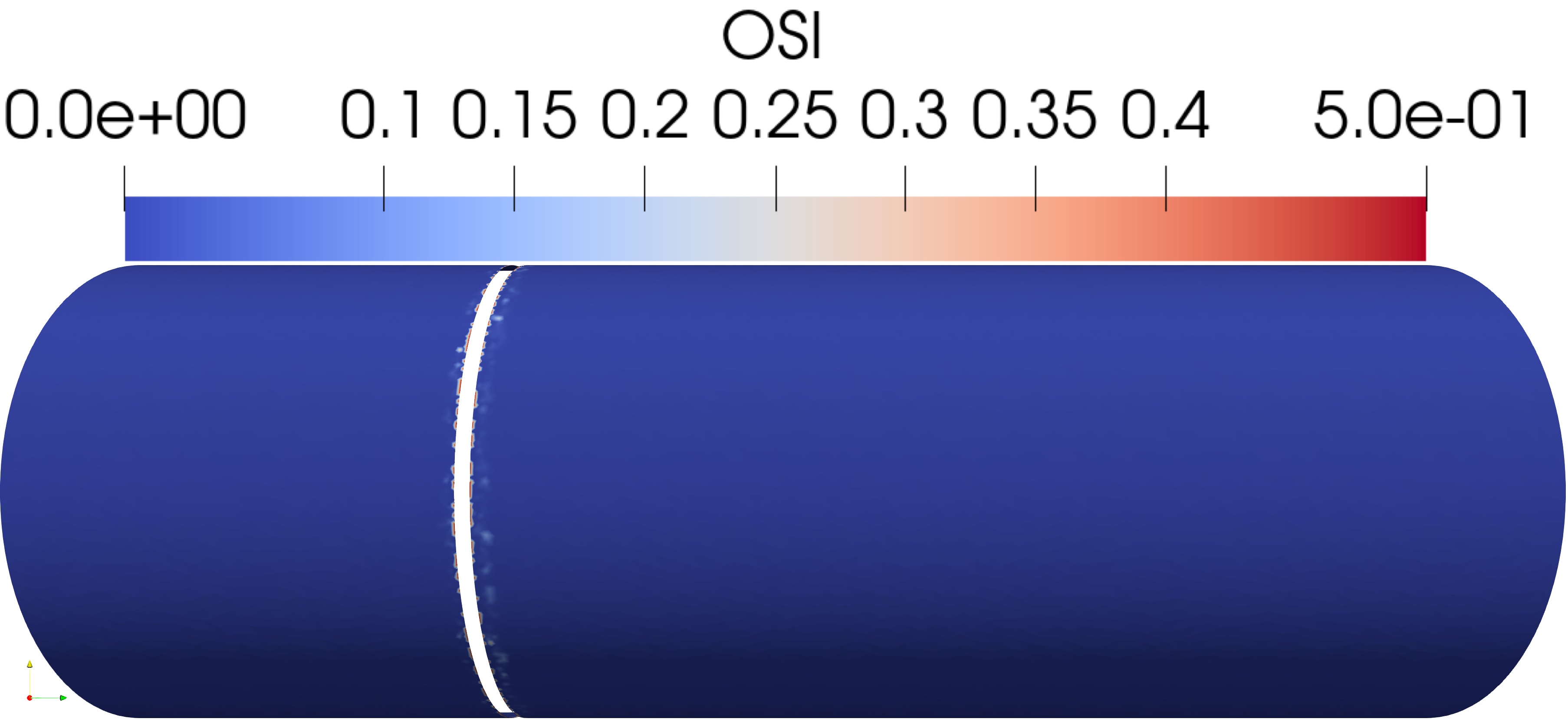}}
    }
    \hfill
    \subfloat[$\rrt_1$]{
    \resizebox*{4.5cm}{!}{\includegraphics[draft=\draftmode]{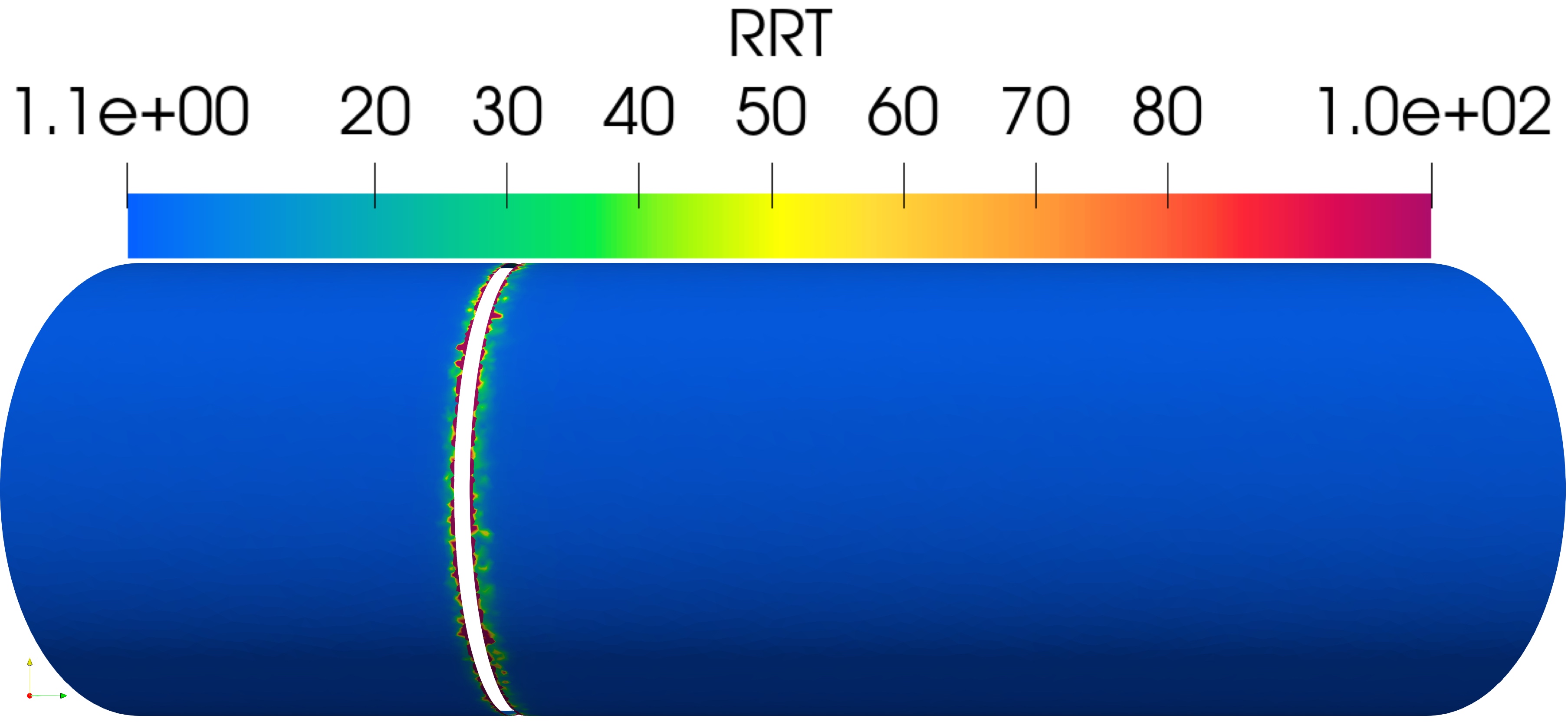}}
    }\\
    \subfloat[$\tawss_2$]{
    \resizebox*{4.5cm}{!}{\includegraphics[draft=\draftmode]{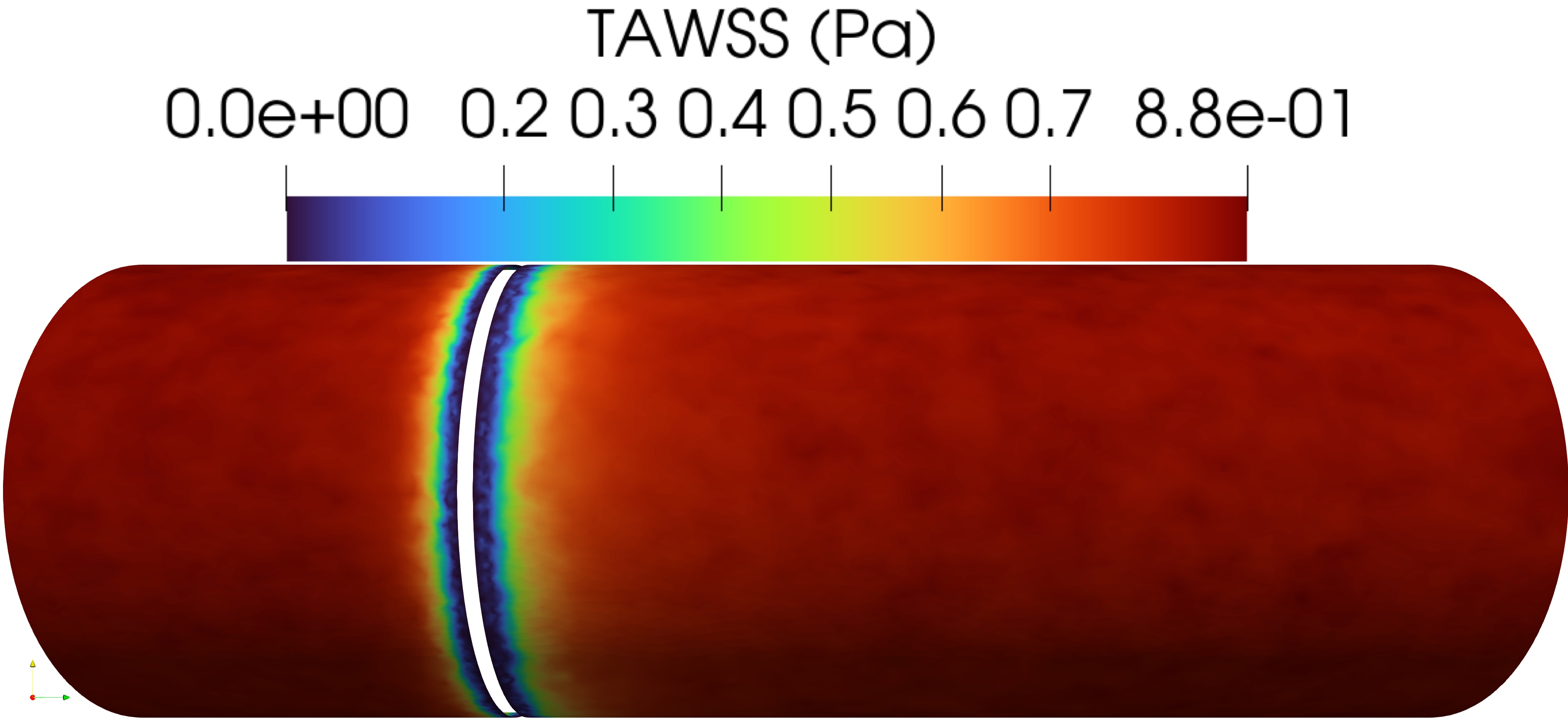}}
    }
    \hfill
    \subfloat[$\osi_2$]{
    \resizebox*{4.5cm}{!}{\includegraphics[draft=\draftmode]{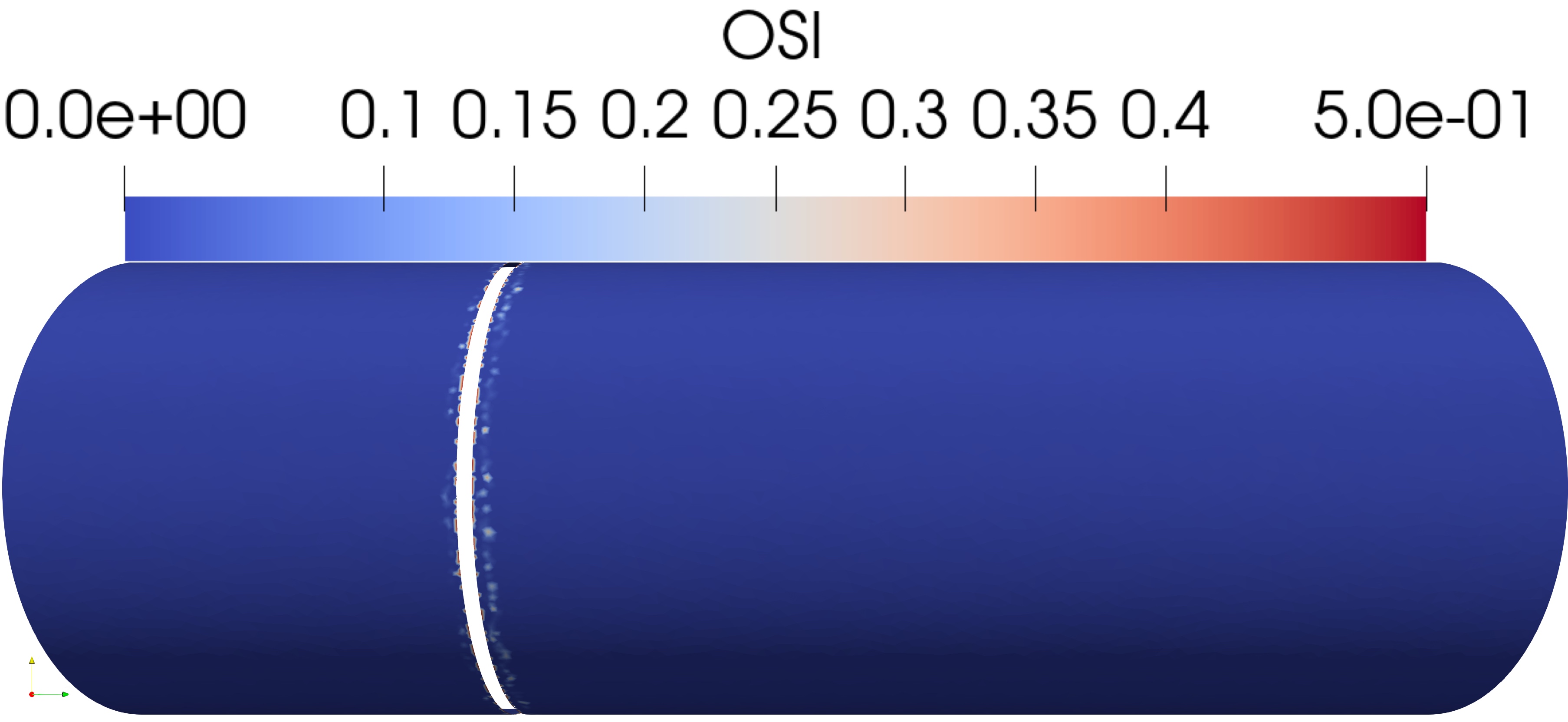}}
    }
    \hfill
    \subfloat[$\rrt_2$]{
    \resizebox*{4.5cm}{!}{\includegraphics[draft=\draftmode]{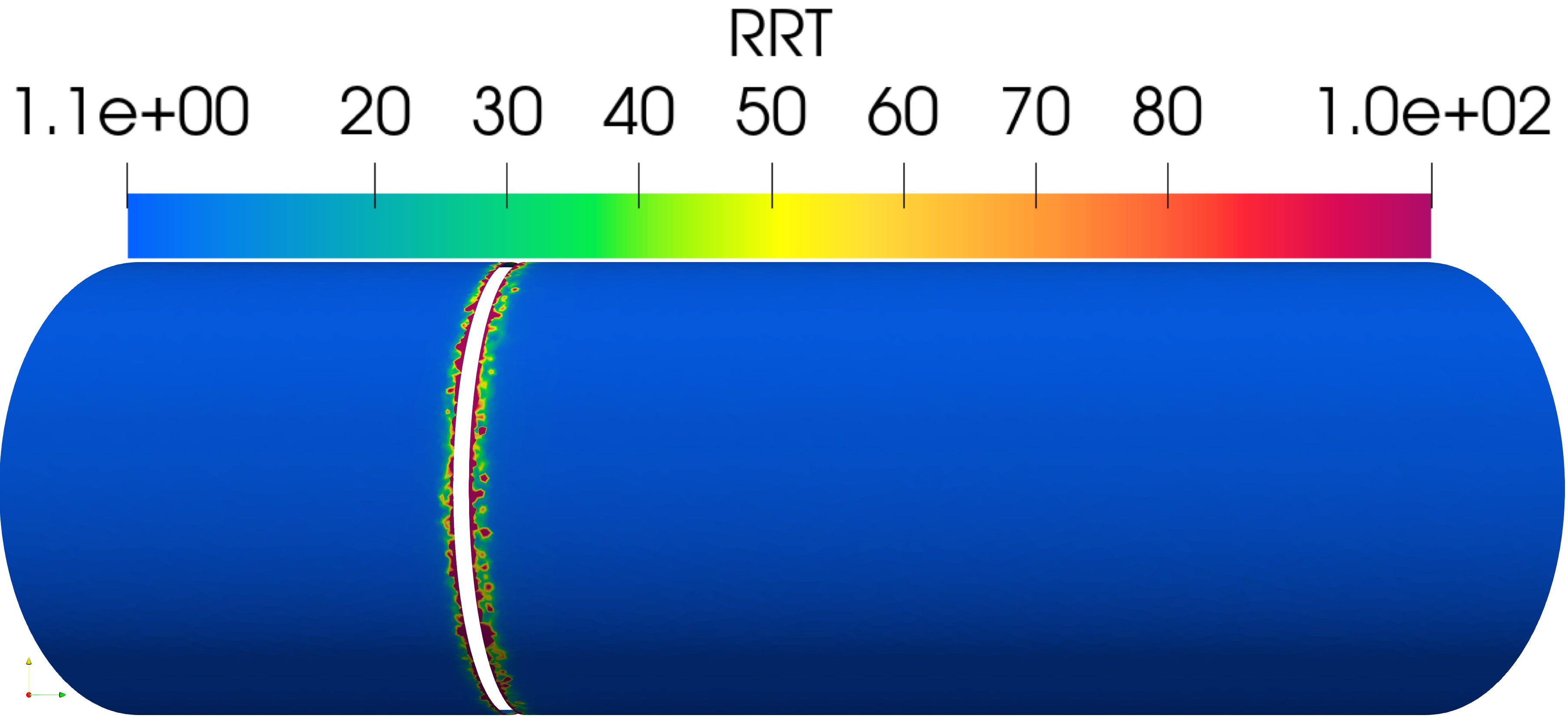}}
    }
    \caption{Artery wall with ring stent.}
    \label{fig:TAWSSOSIRRT-RS}
\end{figure}

Similarly to TAWSS, the overestimation error is further carried on to $\osi_1$ and $\rrt_1$ \review{as shown in Figure \ref{fig:TAWSSOSIRRT-RS}}. In the former case, OSI is underestimated and some recirculation areas are neglected, while in the latter, $\rrt_2$ shows more areas of high RRT, compared to $\rrt_1$.

\begin{figure}[htbp!]
    \subfloat[$\tawss_1$]{
    \resizebox*{1cm}{!}{\includegraphics[draft=\draftmode,trim={35cm 0 85cm 0},clip=true]{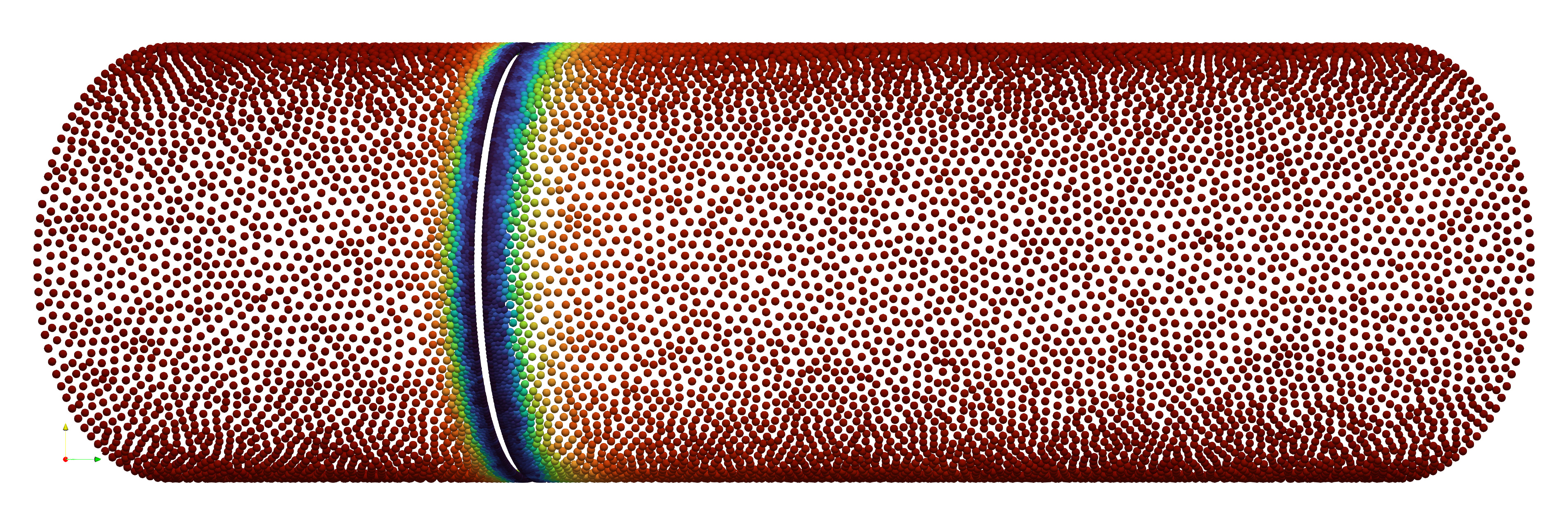}}
    }
    \hfill
    \subfloat[$\tawss_2$]{
    \resizebox*{1cm}{!}{\includegraphics[draft=\draftmode,trim={35cm 0 85cm 0},clip=true]{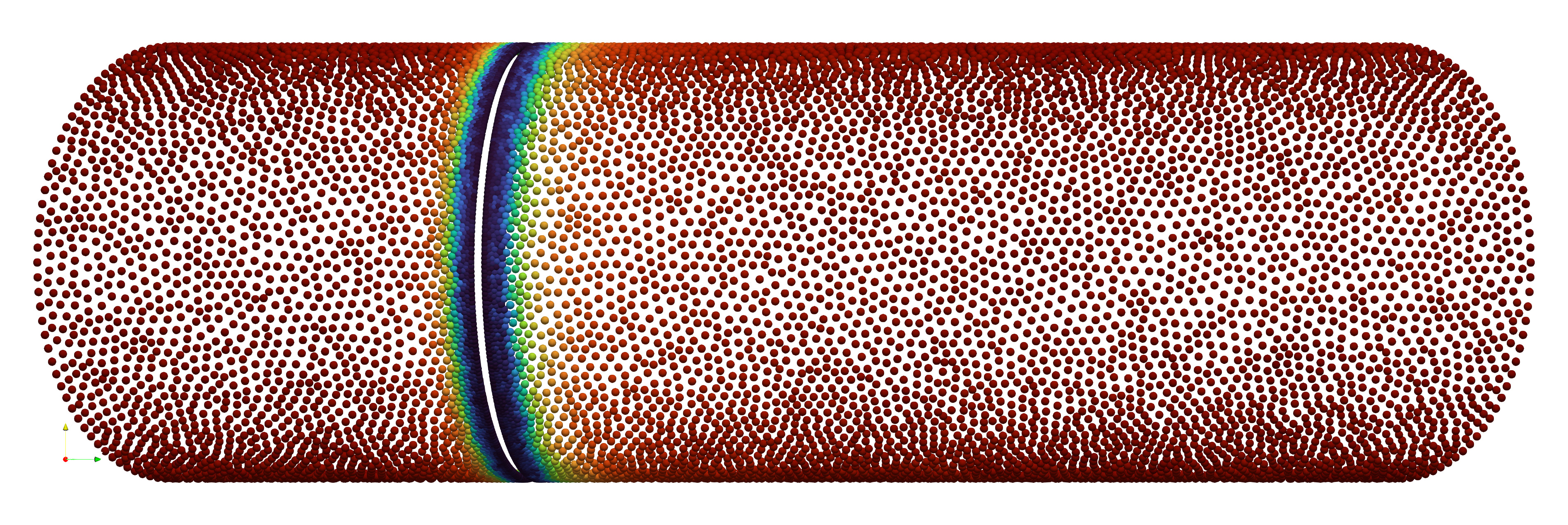}}
    }
    \hfill
    \subfloat[$\epsilon_{\tawss}$]{
    \resizebox*{1cm}{!}{\includegraphics[draft=\draftmode,trim={35cm 0 85cm 0},clip=true]{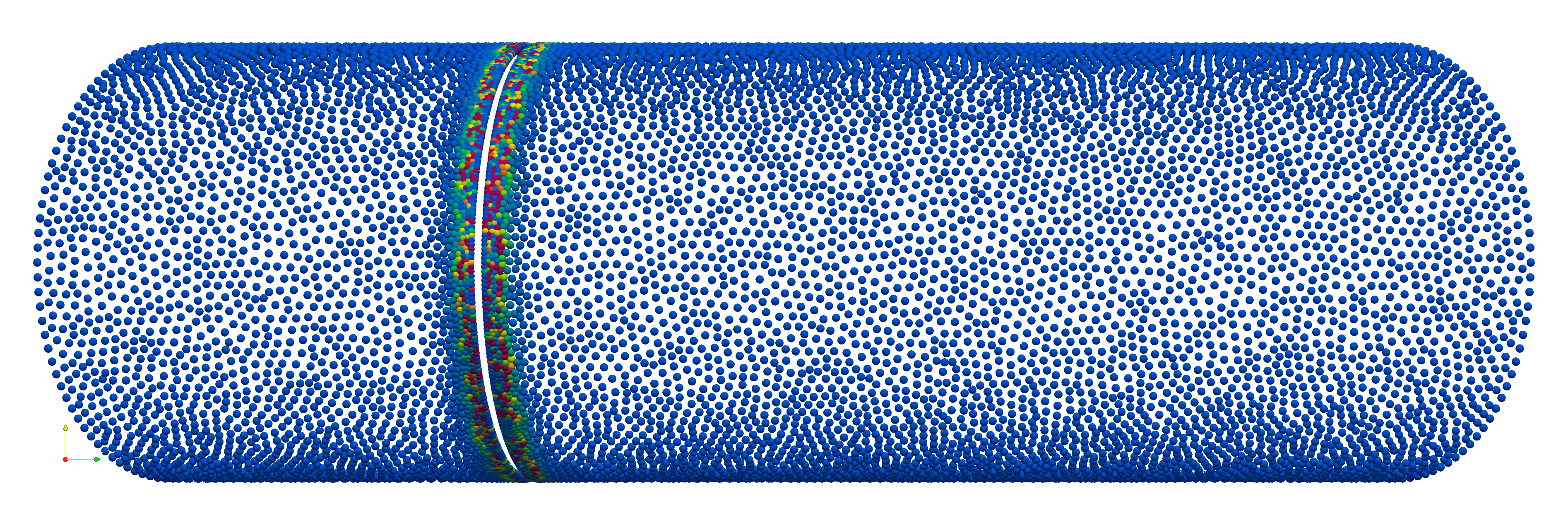}}
    }
    \hfill
    \subfloat[$\delta_1$]{
    \resizebox*{1cm}{!}{\includegraphics[draft=\draftmode,trim={35cm 0 85cm 0},clip=true]{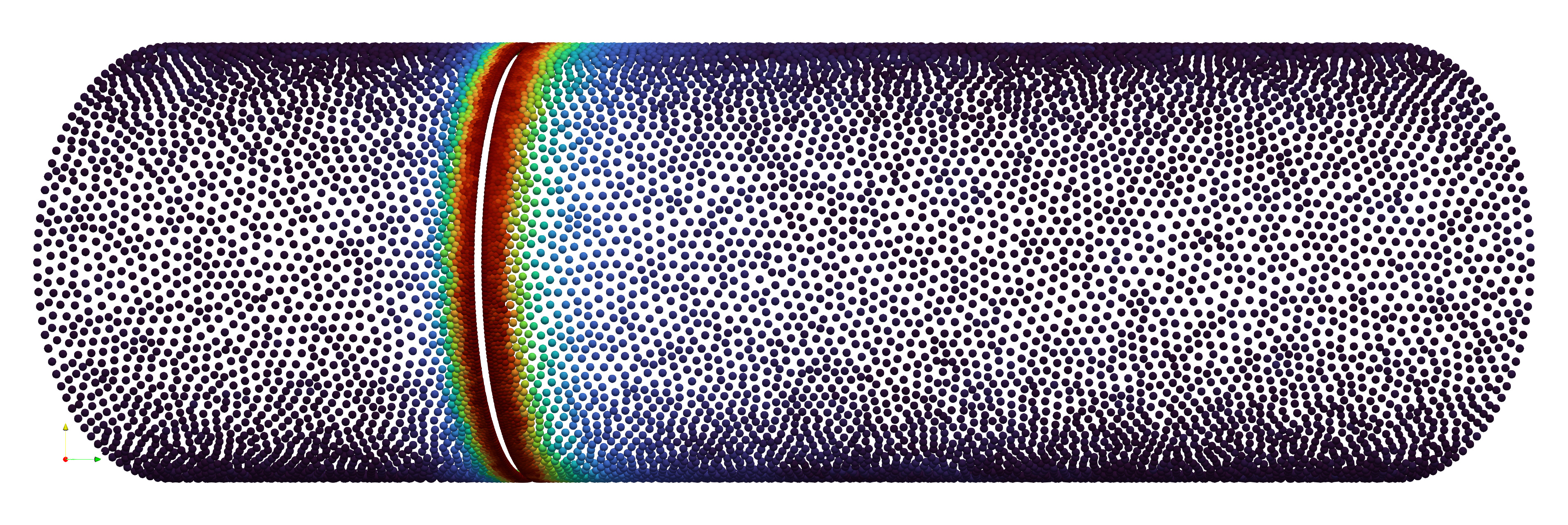}}
    }
    \hfill
    \subfloat[$\delta_2$]{
    \resizebox*{1cm}{!}{\includegraphics[draft=\draftmode,trim={35cm 0 85cm 0},clip=true]{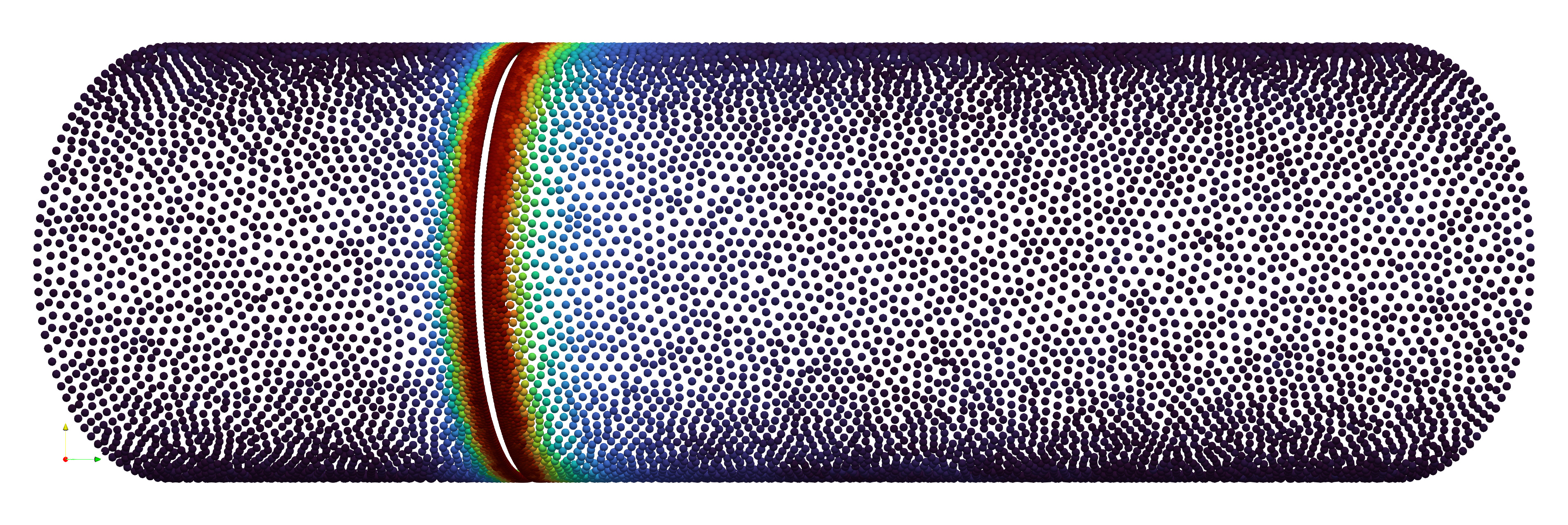}}
    }
    \hfill
    \subfloat[$\osi_1$]{
    \resizebox*{1cm}{!}{\includegraphics[draft=\draftmode,trim={35cm 0 85cm 0},clip=true]{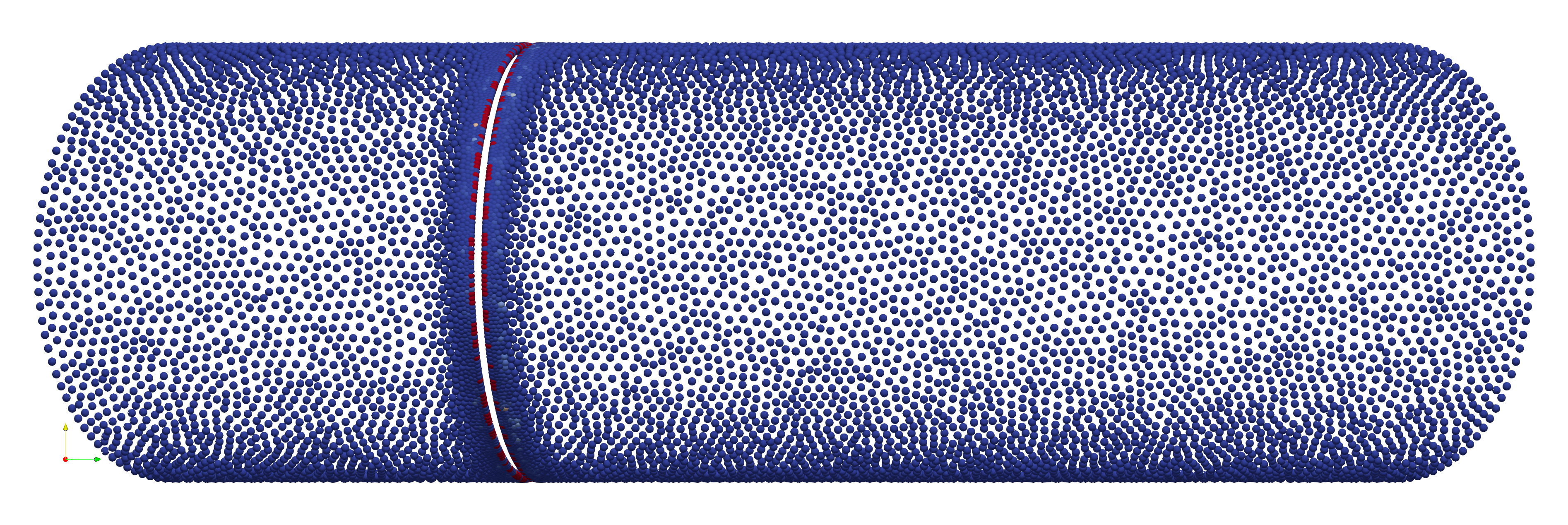}}
    }
    \hfill
    \subfloat[$\osi_2$]{
    \resizebox*{1cm}{!}{\includegraphics[draft=\draftmode,trim={35cm 0 85cm 0},clip=true]{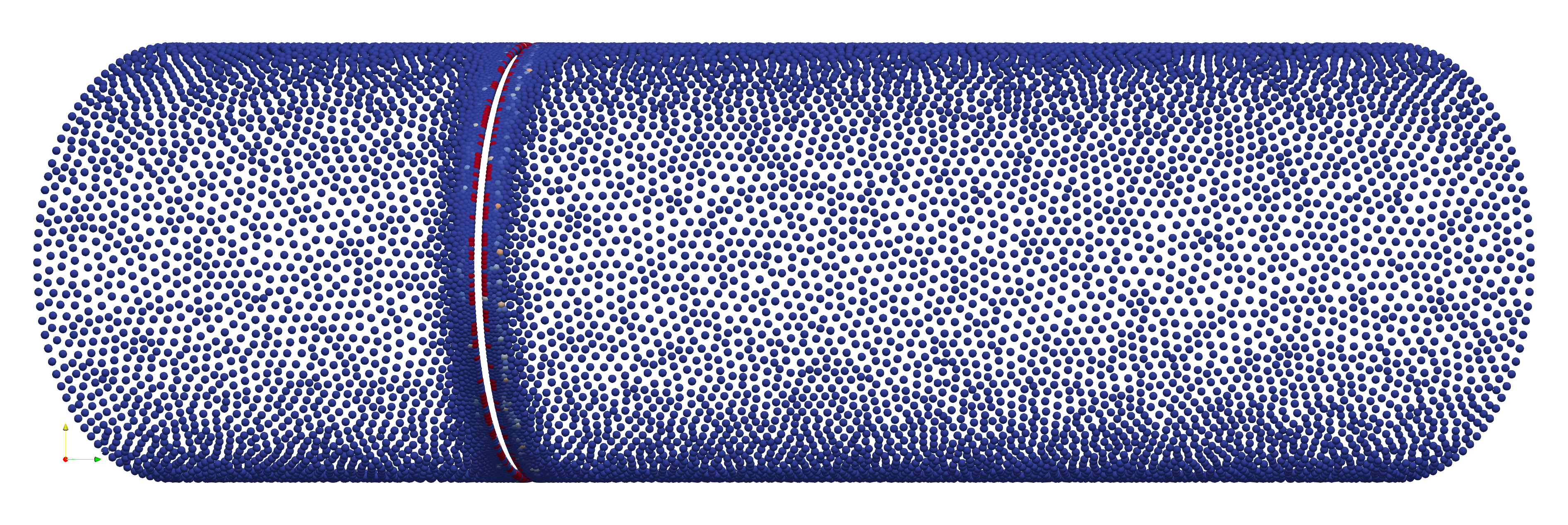}}
    }
    \hfill
    \subfloat[$\rrt_1$]{
    \resizebox*{1cm}{!}{\includegraphics[draft=\draftmode,trim={35cm 0 85cm 0},clip=true]{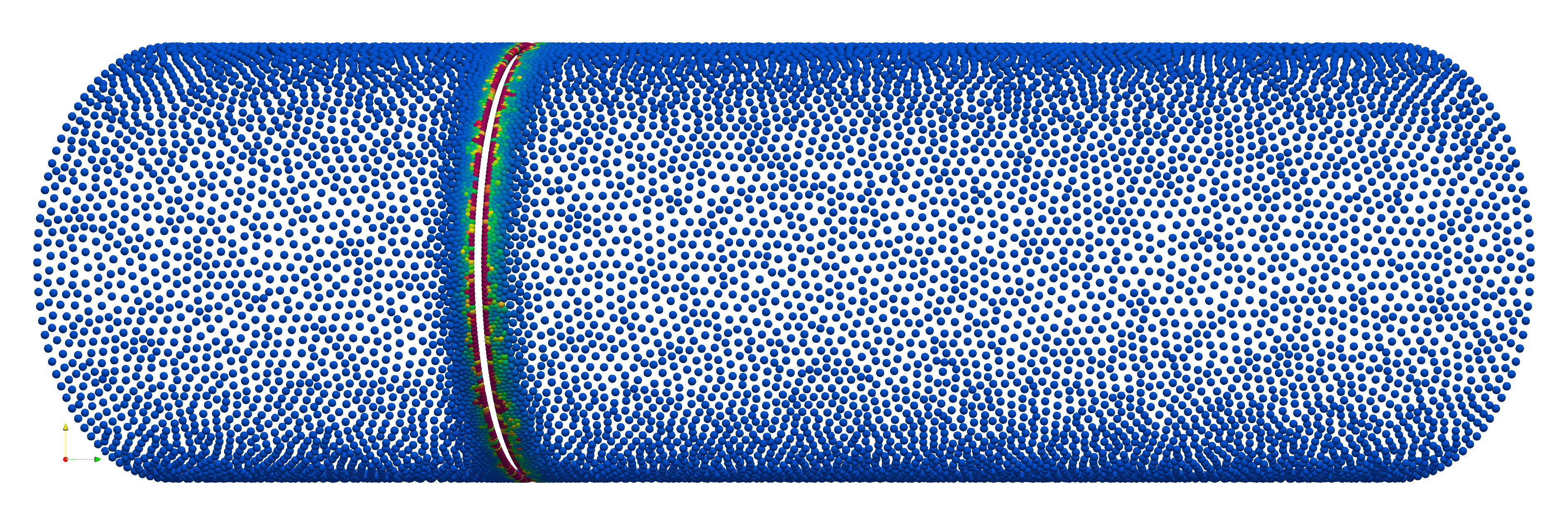}}
    }
    \hfill
    \subfloat[$\rrt_2$]{
    \resizebox*{1cm}{!}{\includegraphics[draft=\draftmode,trim={35cm 0 85cm 0},clip=true]{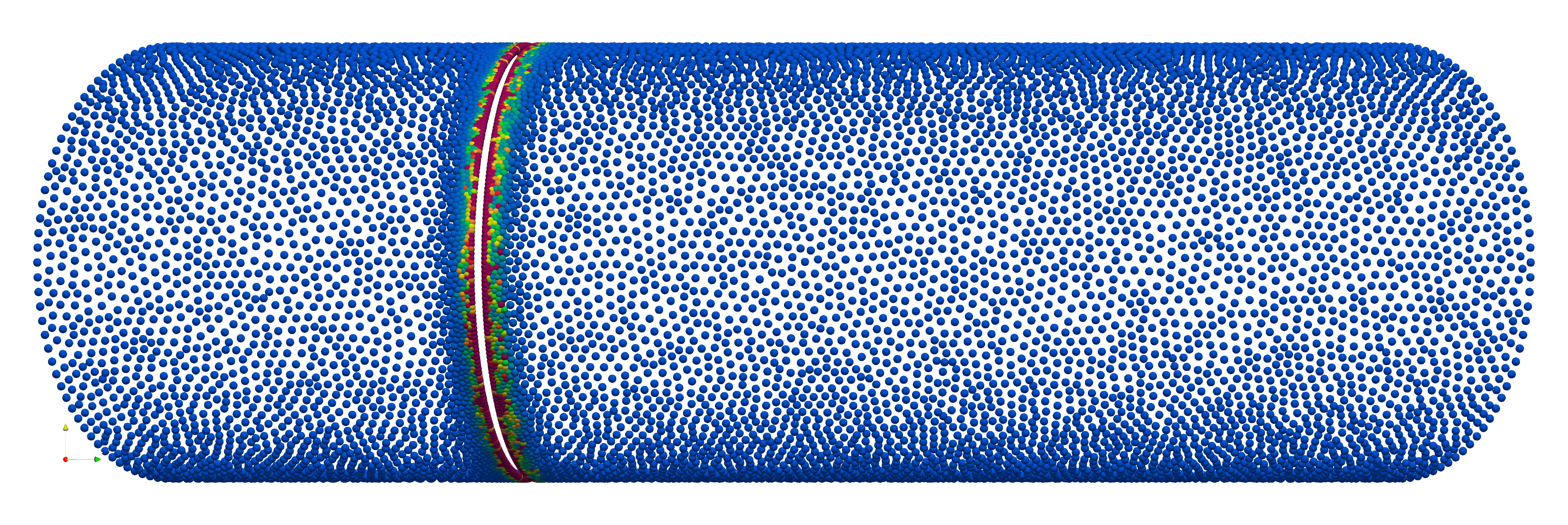}}
    }
    \caption{Zoom on artery wall in stent proximity, see Fig. \ref{fig:WSSStentedWall}.}
    \label{fig:ZoomRS}
\end{figure}

\review{In Figure \ref{fig:ZoomRS}, we focus on the ring stent vicinity. We can qualitatively observe that $\tawss_1$ overestimated the WSS values near the stent and the relative error up to 100\% is carried after time integration from $\wss_1$ onto TAWSS, deviation $\delta$, OSI and RRT}.\\

This analysis concludes that even in the presence of very small struts (two orders of magnitude smaller than the artery length and one order smaller than the radius), the effect of recirculation and non-unidirectional flow cannot be neglected in the stent proximity. The results in the following section will only show values obtained from $\wss_2$, but for simplicity, we drop the subscript.

\subsection{Artery implanted with a XIENCE-V stent in \textit{ad-hoc} configuration}
\begin{figure}[htbp!]
\centering
\subfloat[Comparison of WSS on stented wall.]{%
\resizebox*{!}{5cm}{\includegraphics[draft=\draftmode]{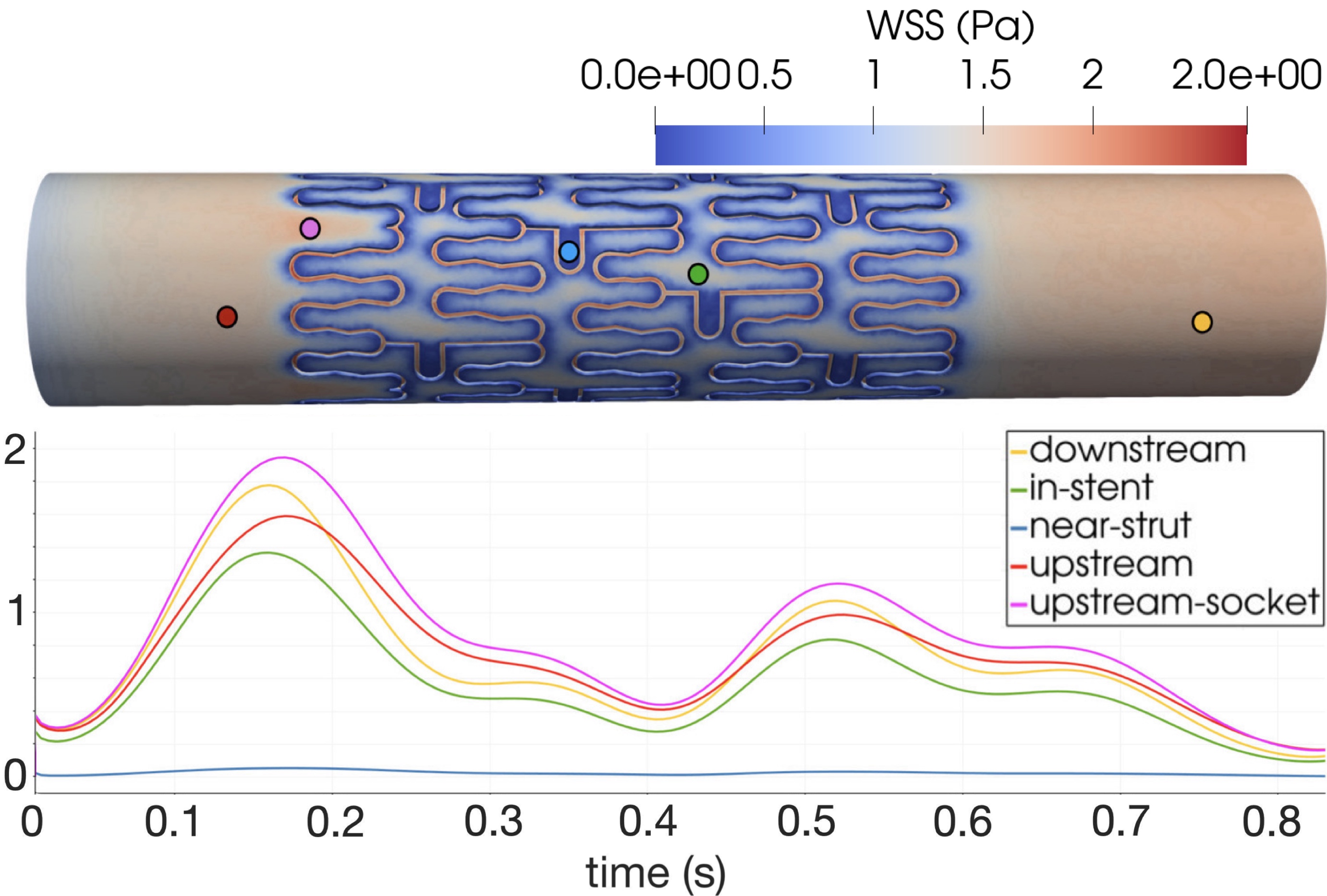}}
\label{fig:UWSS-all}}
\hspace{5pt}
\subfloat[WSS in strut vicinity.]{%
\resizebox*{!}{5cm}{\includegraphics[draft=\draftmode]{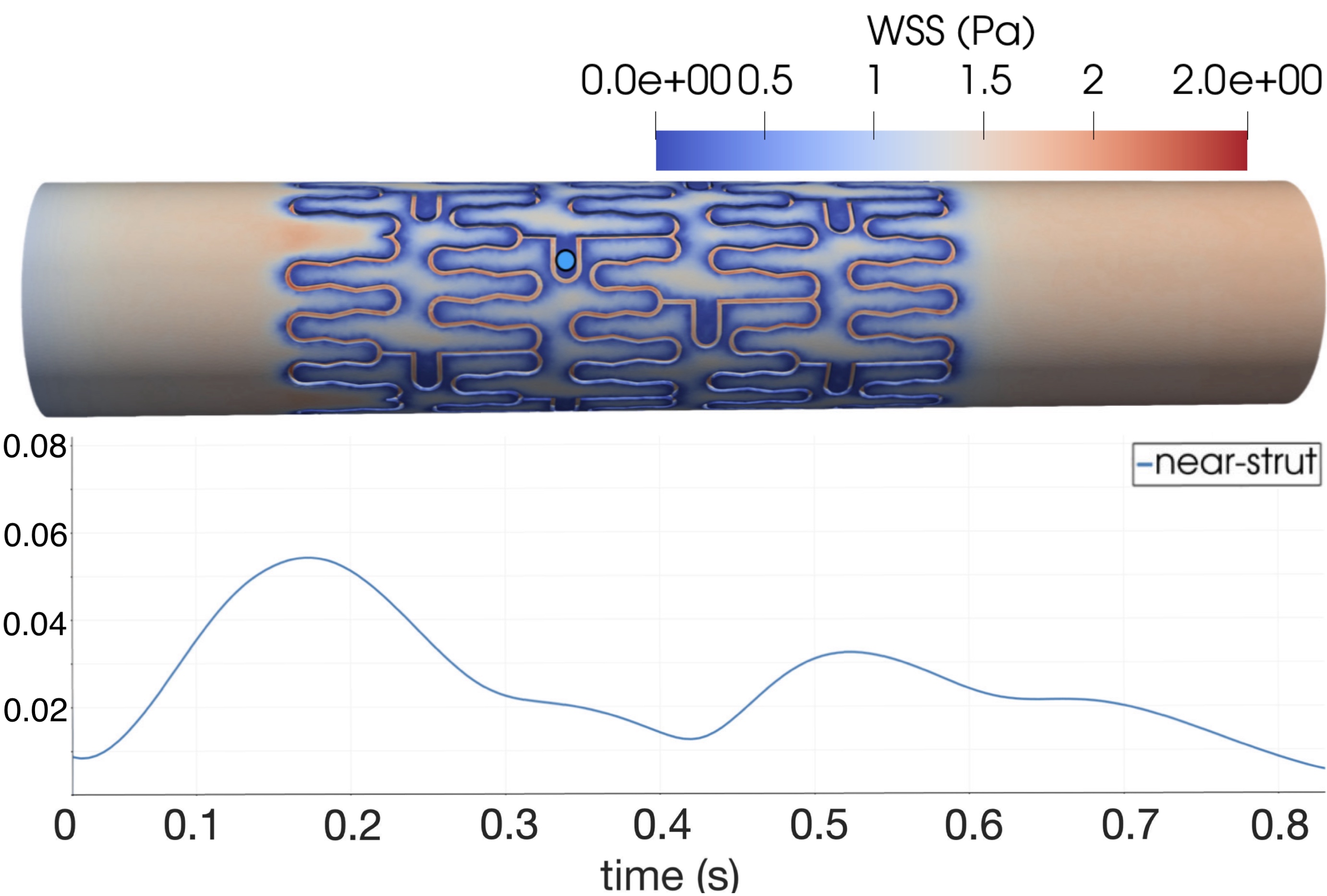}}\label{fig:UWSS-low}}
\caption{Side view of WSS in artery with XIENCE-V stent in ad-hoc configuration at t = 0.25 s (top) and over once cycle T (bottom). Lines are color-coded to the location on the artery wall.}\label{fig:UWSS}
\end{figure}

For a more realistic application case, we simulate hemodynamics in an artery with XIENCE-V stent in expanded ad-hoc configuration and no indentation. The lumen with XIENCE-V stent has circa 2.5 M tetrahedron elements with an average mesh size of 0.07 mm. To better analyze the different areas of WSS, we can look at a still-frame of the stented artery after 0.25 s in Fig. \ref{fig:UWSS-all} where we choose different points on the lumen-wall interface and analyze the WSS. The bottom plot depicts the WSS values of the highlighted areas over one cycle. Starting from the upstream wall, the red dot shows an overall physiological WSS over one cycle, and the pink dot highlights the socket area where the WSS is higher than any other region on the artery wall. The downstream wall (yellow) and the in-stent area (green) show physiological values, as well. Although the latter is lower than the average, it is still considered to be in an acceptable range. Extremely low values of WSS are detected near the struts, in particular within the U-shaped struts highlighted by the blue dot. The WSS values shown in Fig. \ref{fig:UWSS-low} are in the range of $10^{-2}$, between 0.01 and 0.08 Pa, which is one order of magnitude lower than the healthy physiological range.

\section{Conclusions and outlook}

A high-fidelity multiphysics framework is herein presented that captures the intricate mechanism of ISR by modeling the significant mediators via suitable governing equations. The influence of the pharmacokinetics and pharmacodynamics in the vessel wall is investigated, which indicates the scope for optimization of drug embedment in modern DESs patient-specifically. In addition, the blood flow model is validated using an idealized healthy artery, approximated as a perfect cylinder, and the hemodynamics are compared to a ring stent test case. Particular emphasis is given to the influence of stent struts on micro dynamics: the analysis of streamlines in section \ref{ResultsRS} shows that recirculation areas and vortices are located near the stent struts and that their geometry strongly influences them. Furthermore, two approaches for the computation of WSS and its derived indicators are compared, concluding that $\wss_2$ is a better measure. Critically low values of WSS are detected near the struts where inflammation and ISR occur, also in the application case with XIENCE-V stent. Our analysis concludes that the mesh size of lumen computational domains in stented arteries has to be at least comparable to the stent thickness to ensure enough accuracy to capture local microdynamics. To reduce computational costs, targeted mesh refinement in the struts vicinity should be considered.

Future work shall then be focused on the coupling between the multiphysics arterial wall model and the hemodynamics model to exchange key indicators that throttle the restenotic response of arteries after stent implantation. Also, improvement in the computational efficiency of the high-fidelity models is to be achieved via the usage of reduced-integration finite elements and model-order reduction techniques.

\section{Acknowledgements}
The financial support of the Deutsche Forschungsgemeinschaft (DFG, German Research Foundation) for the projects ``Drug-eluting coronary stents in stenosed arteries: medical investigation and computational modelling'' (project number 395712048: RE 1057/44-1, RE 1057/44-2), ``In-stent restenosis in coronary arteries - in silico investigations based on patient-specific data and meta modeling" (project number 465213526: RE 1057/53-1), a subproject of ``SPP 2311:  Robust coupling of continuum-biomechanical in silico models to establish active biological system models for later use in clinical applications - Co-design of modeling, numerics and usability", and ``Modelling of Structure and Fluid-Structure Interaction during Tissue Maturation in Biohybrid Heart Valves'', a subproject of ‘‘PAK961 - Modeling of the structure and fluid–structure interaction of biohybrid heart valves on tissue maturation’’ (project number 403471716: RE 1057/45-1, RE 1057/45-2) is gratefully acknowledged. Additionally, this work was partially supported by the DFG via the project ``GRK 2379:  Hierarchical and Hybrid Approaches in Modern Inverse Problems" (project number 333849990).

The authors also gratefully acknowledge the computing time granted by the JARA Vergabegremium and provided on the JARA Partition part of the supercomputer CLAIX at RWTH Aachen University and on the supercomputer JURECA at Forschungszentrum Jülich. In addition, the authors gratefully acknowledge the computing time provided to them on the high-performance computer Lichtenberg at the NHR Centers NHR4CES at TU Darmstadt. This is funded by the Federal Ministry of Education and Research, and the state governments participating on the basis of the resolutions of the GWK for national high-performance computing at universities (www.nhr-verein.de/unsere-partner).

\bibliography{GAMM_Mitteilungen}

\end{document}